\documentclass[aps,prb]{revtex4}
\usepackage{graphicx}
\usepackage{bm}
\usepackage{xfrac}
\usepackage{amsmath}
\usepackage{amssymb}
\usepackage{bbold}
\usepackage{color}
\usepackage{tikz}
\usepackage{xcolor}

\newcommand{\be}{\begin{equation}}
\newcommand{\ee}{\end{equation}}
\newcommand{\bea}{\begin{eqnarray}}
\newcommand{\eea}{\end{eqnarray}}
\newcommand{\beb}{\begin{eqnarray*}}
\newcommand{\eeb}{\end{eqnarray*}}

\newcommand{\Tr}{\text{Tr}}
\newcommand{\Hh}{\mathcal{H}}

\newcommand{\cc}{\text{c}}
 
\newcommand{\h}{\text{h}}
\begin{document}
\title{Edge structure of graphene monolayers in the $\nu=0$ quantum Hall state}

\author{Angelika Knothe$^{1,2}$}
\author{Thierry Jolicoeur$^{1}$}
\affiliation{1) Laboratoire de Physique Th\'eorique et Mod\`eles statistiques,
Universit\'e Paris-Sud, 91405 Orsay, France\\
2) Physikalisches Institut, Albert-Ludwigs-Universit\"at Freiburg, 
Hermann-Herder-Str. 3, D-79104 Freiburg, Germany}

\date{July, 2015}
\begin{abstract}
Monolayer graphene at neutrality in the quantum Hall regime has many competing
ground states with various types of ordering.
The outcome of this competition is modified by the presence of the sample boundaries.
In this paper we use a Hartree-Fock treatment of the electronic correlations allowing
for space-dependent ordering. The armchair edge influence is modeled by a simple perturbative
effective magnetic field in valley space.
We find that all phases found in the bulk of the sample, ferromagnetic, canted antiferromagnetic,
charge-density wave and Kekul\'e distortion are smoothly connected to a Kekul\'e-distorted
edge. The single-particle excitations are computed taking into account the spatial variation
of the order parameters. 
An eventual metal-insulator transition as a function of the Zeeman energy is not simply
related to the type of bulk order.
\end{abstract}
\maketitle

\section{introduction}

When subjected to a strong perpendicular magnetic field, the 
electrons confined to the two-dimensional (2D) carbon lattice of graphene form a 
unique quantum Hall (QH) system. Notably graphene at neutrality is an example
of QH ferromagnetism with many competing ground 
states~\cite{MacDonald06, Herbut1, Herbut2, Alicea,yang_collective_2006,jung_theory_2009,
kharitonov_phase_2012, abanin_fractional_2013, sodemann_broken_2014}. 
In such systems
we expect generically complex physics at the edge.
Early work on graphene edge states~\cite{abanin_spin-filtered_2006} has shown that
when taking into account the electron spin degree of freedom, the edge states 
should be of helical nature, i.e., exhibiting counterpropagating modes 
carrying opposite spin polarizations. Graphene was thus proposed to be a model 
candidate for a quantum spin Hall (QSH) system.
Furthermore, graphene has unique 
features so that it may be an ideal probe material to 
study QH edge physics experimentally~: in semiconductor-based 2D electron 
gas systems the confining potential is soft at the scale of the magnetic length so that
there may be 
 edge reconstruction~\cite{chklovskii_electrostatics_1992, 
chamon_sharp_1994, yang_field_2003}, spoiling the ideal theoretical description. Therefore, the 
understanding of edge phenomena in these 2D systems 
is difficult~\cite{chang_observation_1996, 
grayson_continuum_1998} and still controversial.
In contrast, the boundaries of the graphene lattice are 
directly defined at the atomic level. Therefore, they naturally represent atomically sharp QH
edges. This should allow  observation of QH edge state 
physics without complications from edge reconstruction~\cite{Bhatt,li_evolution_2013}. The 
fabrication, design, and control of graphene edge structures with atomic level precision 
 is a field of ongoing research~: see, e.g., Refs.~\onlinecite{fuhrer_graphene:_2010, cai_atomically_2010, huang_spatially_2012}. 
Among the theoretical approaches, also a mean-field treatment of a Hubbard-type model has been applied to the edge physics~\cite{Lado_noncollinear_2014}.
The QSH nature of the graphene edge has been the subject of recent 
experimental investigations~\cite{young_tunable_2014}.
 At Landau level filling 
factor $\nu=0$, upon tilting the applied magnetic field  with respect to the 
graphene sheet, there is a metal-insulator transition for
some critical angle. This suggests a change of  
the bulk state as a function of tilting. Indeed, extensive theoretical studies 
of the $\nu=0$ ground state (GS) structure of bulk 
graphene~\cite{jung_theory_2009,kharitonov_phase_2012,roy1,roy2}
have shown the existence of various competing phases with distinct symmetry-breaking  properties. 
While the graphene GS at neutrality is a highly degenerate SU(4) ferromagnetic multiplet,
small symmetry-breaking terms due to short-distance physics lift this degeneracy
and the system may form various different phases. Among these phases are e.g. 
the ferromagnet (F) state\cite{MacDonald06, yang_collective_2006} or the antiferromagnet (AF) state, 
where the latter may form a canted antiferromagnetic (CAF) state as has been pointed 
out by Kharitonov in Ref.~\onlinecite{kharitonov_phase_2012}. 
Further possible phases include 
a charge-density wave\cite{Alicea, Alicea2, Fuchs, jung_theory_2009} (CDW) or a Kekul\'e distorted state\cite{Nomura2, Hou} (KD).
Transitions between these states may be induced by varying the Zeeman energy
which is done by tilting the field.

In this paper we study the edge structure of the $\nu=0$ QH state
 in the presence of SU(4)-symmetry-breaking interactions. 
We use a simple model of the edge potential in the basis of the $n=0$ LL orbitals
appropriate to an armchair termination of the graphene lattice
and treat interactions by a Hartree-Fock (HF) approximation. Our variational ansatz
is orbital-dependent so it captures the spatial variations of the ordering from the
bulk to the edge (an effect which is absent in previous HF studies~\cite{kharitonov_edge_2012}).
We find that there is always a crossover towards a Kekul\'e distorted region close to the edge.
There is appearance also of spin/valley nontrivial entanglement which is limited
to the transition region towards Kekul\'e order and does not take place either in the bulk
or close to the edge.
We propose a quantitative measurement of the entanglement by computing the concurrence
as a function of edge distance.
Within HF we also compute the single-particle properties of the particle-hole excitation spectrum.
We discuss how this spectrum varies with the edge distance and also how it is influenced by the nature
of the bulk order. Our main finding is that there should be always a metal-insulator transition
as a function of the Zeeman field whatever the nature of the ordered phase. So strictly speaking
 the experimental observations of Ref.~\onlinecite{young_tunable_2014} do not imply that 
their graphene bulk is a CAF state.
It should be noted that our HF treatment has some shortcomings. Notably the Coulomb exchange interaction
in the $n=0$ LL leads to a coupling between charge and spin/valley degrees of freedom. So in general
charge motion is done through spin/valley textures as proposed in 
Refs.~\onlinecite{brey_electronic_2006,brey_edge_2006,fertig_luttinger_2006,murthy_collective_2014}.
As in previous HF works~\cite{kharitonov_edge_2012} we will not try to model these effects.
While the edge effects will overcome exchange energy close enough to boundary, they may change
the nature of excitations right in the transition region. More work is needed to understand this point.

The paper is structured as follows~: In Sec.~\ref{section:TheoFrame}, we 
define the theoretical framework describing the $\nu=0$ QH state of
graphene in the presence of a boundary and the corresponding 
model Hamiltonian. We introduce the 
parametrization of the Hartree-Fock (HF) GS wave function in Sec.~\ref{section:HF_treat}.
In Sec.~\ref{section:PropGS}, we present 
 results for the GS and its properties obtained from minimizing the HF energy 
functional. We describe the evolution of the different possible bulk phases 
when moving  towards to the edge. We find that  the presence of a boundary
gives rise to novel spin/isospin configurations which do not exist in the
bulk. Furthermore, we show the existence of an intermediate region between the bulk and 
the edge with nontrivial entanglement of spin and valley isospin. 
In Sec.~\ref{ssection:MFSpectrum}, we 
extend our HF treatment to compute the spatial evolution of the single particle (SP) energy levels and 
corresponding eigenstates from the bulk to the boundary. As pointed out by 
previous work~\citep{kharitonov_edge_2012}, the SP spectra can either exhibit 
nonzero gaps or support gapless edge states, depending on the system 
parameters determining the bulk phase. We describe how the spatial variation 
of the spin/isospin texture influences 
the shape of the SP energy levels. This leads to an understanding of  the edge gap, the 
number of conducting channels, as well as possible conclusions for the 
bulk symmetry properties  drawn from the conductance behavior of 
the edges. 
In a final part, we compute the spin and isospin properties of the 
corresponding SP eigenstates to directly prove
that the edge states of the $\nu=0$ QH state in graphene indeed exhibit the 
helical properties of a QSH state. 
Sec.~\ref{section:Discussion} finally 
discusses our results in relation to experimental findings and contains our conclusions.

\section{Model of the Graphene edge}
\label{section:TheoFrame}
We first recall basic facts about the electronic structure of 
graphene~\cite{castro_neto_electronic_2009,goerbig_electronic_2011}.
The
hexagonal structure admits two triangular Bravais sublattices $A$ and 
$B$  that form the basis for a  tight-binding Hamiltonian. In the Brillouin zone
there are two special degeneracy points~:
at these Dirac points $K$ and $K^\prime$, 
the valence and the conductance band form Dirac cones
and touch at the Fermi level for neutral graphene.
In the vicinity of the Dirac points, for each orientation of the spin $\sigma=\uparrow,\downarrow,$ the
electronic wave functions $\Psi_{A}$ and $\Psi_{B}$ on the two sublattices can be written as~:
\begin{subequations}
\begin{align}
\Psi_{A {,\sigma}}(\mathbf{r}) &= e^{i\mathbf{K}\cdot \mathbf{r}} \psi_{A  {,\sigma}} +
e^{-i\mathbf{K}^{\prime}\cdot \mathbf{r}} \psi_{A  {,\sigma}}^{\prime},\\
\Psi_{B{,\sigma}}(\mathbf{r}) &= e^{i\mathbf{K}\cdot \mathbf{r}} \psi_{B {,\sigma}} +
e^{-i\mathbf{K}^{\prime}\cdot \mathbf{r}} \psi_{B  {,\sigma}}^{\prime}.
\end{align}
\label{eqn:PsiAB}
\end{subequations}
Assembling the envelope functions in a four spinor notation we write for the 
electronic state~:
\begin{equation}\Psi_{ {\sigma}}=
\begin{pmatrix} 
\psi_{A {,\sigma}}\\
\psi_{B {,\sigma}}\\
\psi_{A {,\sigma}}^{\prime}\\
\psi_{B{,\sigma}}^{\prime}
\end{pmatrix} _{{ H_{K,K^{\prime}} \otimes H_{A,B} } },
\label{eqn:Psi}
\end{equation}
{where the subindex $H_{K,K^{\prime}} \otimes H_{A,B} $ indicates that the state $\Psi_{\sigma}$ lives 
in the Hilbert space formed as the direct product between Dirac valley space $H_{K,K^{\prime}}$ and the $A,B$ sublattice space $H_{A,B}$.}
An applied magnetic field leads to the formation of  Landau 
levels (LL) with energies~:
\begin{equation}
E_{n}=\frac{v_F}{\ell_B}\sqrt{2|n|}\;\text{sgn(}{n}\text{)},
\label{eqn:LLevel}
\end{equation}
where $n\in\mathbb{Z}$ is the LL index, $\ell_B=\sqrt{\frac{\hbar c}{e 
B_{\perp}}}$ is the magnetic length for  
 perpendicular magnetic field strength $B_{\perp}$, and $v_F$ denotes the 
Fermi velocity. The corresponding eigenstates can be written as~:
\begin{equation}\Psi_{n\neq0 {,\sigma}}=\frac{1}{\sqrt{2}}
\begin{pmatrix} 
|n-1\rangle\\
|n\rangle\\
-|n-1\rangle\\
-|n\rangle
\end{pmatrix}_{{ H_{K,K^{\prime}} \otimes H_{A,B} }} , \hskip10pt \Psi_{n=0 {,\sigma}}=
\begin{pmatrix} 
0\\
|0\rangle\\
0\\
-|0\rangle
\end{pmatrix}_{{ H_{K,K^{\prime}} \otimes H_{A,B} }} .
\label{eqn:Psi_n}
\end{equation}
The filling factor $\nu$ of the Landau levels is defined by~:
\begin{equation}
\nu=\frac{n_e}{2\pi\ell^2_B},
\label{eqn:Def_nu}
\end{equation}
where $n_e$ denotes the electronic density. The configuration of 
neutral graphene, i.e., $\nu=0$ is peculiar. Indeed this 
particle-hole symmetric situation corresponds to the case in which all the LLs 
with $n<0$ are filled and all the LLs with $n>0$ are empty, but the $n=0$ LL is 
exactly half filled with two electrons per orbital. The form of the 
$n=0$ wave function as given in Eq.~(\ref{eqn:Psi_n})  implies that 
$n=0$ LL electrons reside on one of the sub-lattices only. 
In the following, we consider the case of neutral graphene and therefore 
study the properties of electrons in the $n=0$  LL. We simplify the 
notation by collecting only the non-zero entries of the $n=0$ spinor in 
Eq.~(\ref{eqn:Psi_n}) as~:
\begin{equation} \Psi_0=
\begin{pmatrix} 
|\uparrow+\rangle\\
|\uparrow-\rangle\\
|\downarrow+\rangle\\
|\downarrow-\rangle
\end{pmatrix} _{{H}},
\label{eqn:Psi_n0}
\end{equation}
identifying the valley and the sublattice indices in a common valley-isospin 
$\tau$ as $\tau=+\mathrel{\hat=}K\leftrightarrow A$ and 
$\tau=-\mathrel{\hat=}K^{\prime}\leftrightarrow B$. In the four-dimensional 
Hilbert space $H=H_{spin}\otimes H_{valley}$ we use the indices $\mu,\nu$ to 
label the four possible configurations of spin and isospin in this space: 
$\mu,\nu\in\{ \uparrow+,\uparrow-, \downarrow+, \downarrow- \}$. {Due to the fourfold degeneracy in spin space ($\sigma=\uparrow, 
\downarrow$) and in valley space ($\tau=+,-$), integer QH effect in graphene is expected at 
values of $\nu$ that change in steps of four.}

\begin{figure}[t]
  \centering
\includegraphics[width=0.6\textwidth]{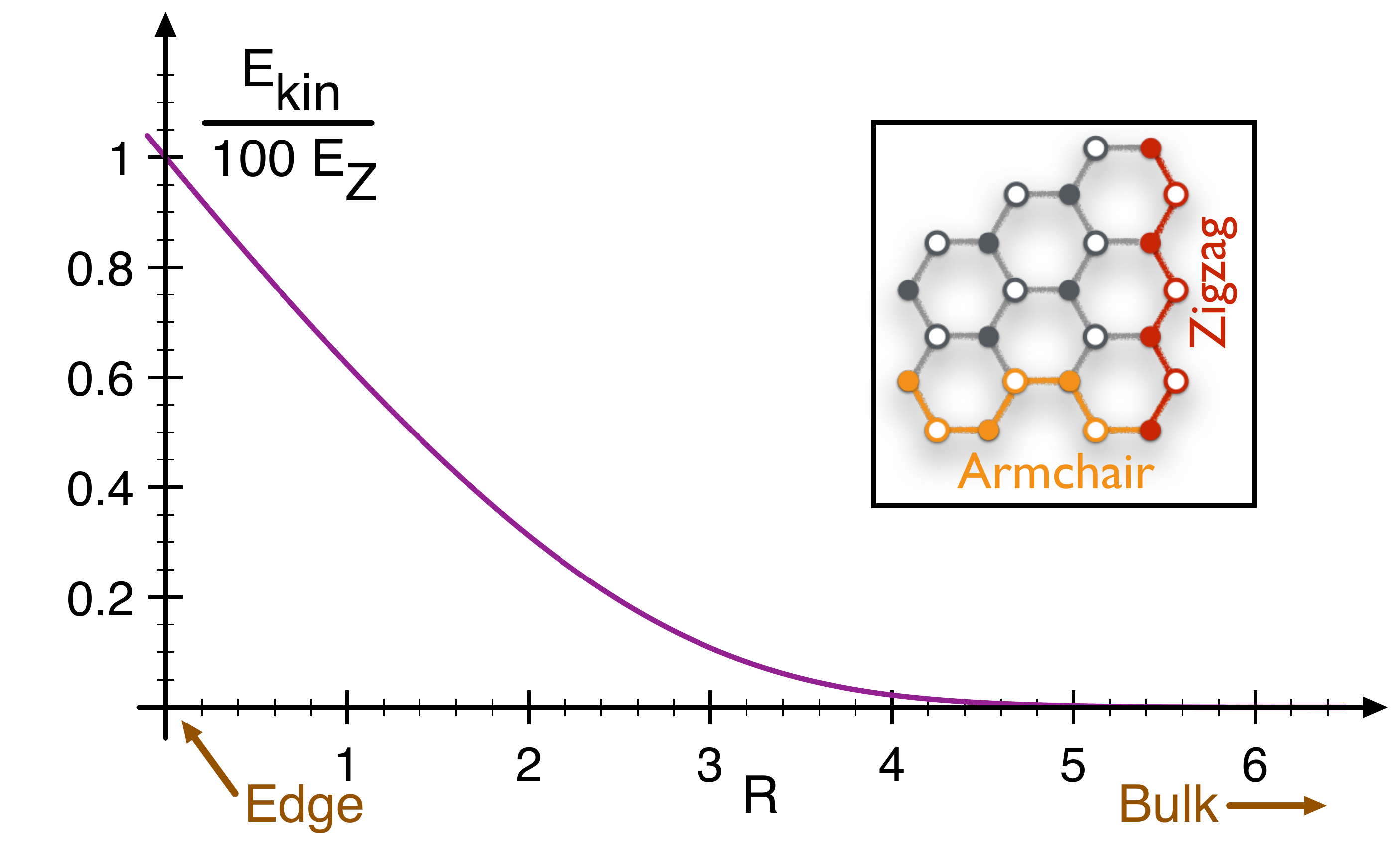}
  \caption{Shape of the kinetic energy edge potential $E_{kin}$ in $\Hh_{kin}$ 
of Eq.~(\ref{eqn:Hkin}). The kinetic energy rises from $E_{kin}=0$ in the bulk 
(equivalent to the case of infinitely extended, translationally invariant 
graphene in the $n=0$ LL) to the energy of the $n=1$ LL exactly at the edge. 
The curve was obtained from solving the 
problem of free electrons on a graphene lattice in the presence of a magnetic 
field in the presence of boundary, i.e., by applying appropriate 
boundary conditions for an armchair edge (see inset).
Zigzag edges have similar nonzero energy states but also additional zero-energy modes
that are beoynd the scope of our work.}
\label{fig:Ekin_vsX_wGrLatt}
\end{figure}

\begin{figure}[t]
  \centering
\includegraphics[width=0.6\textwidth]{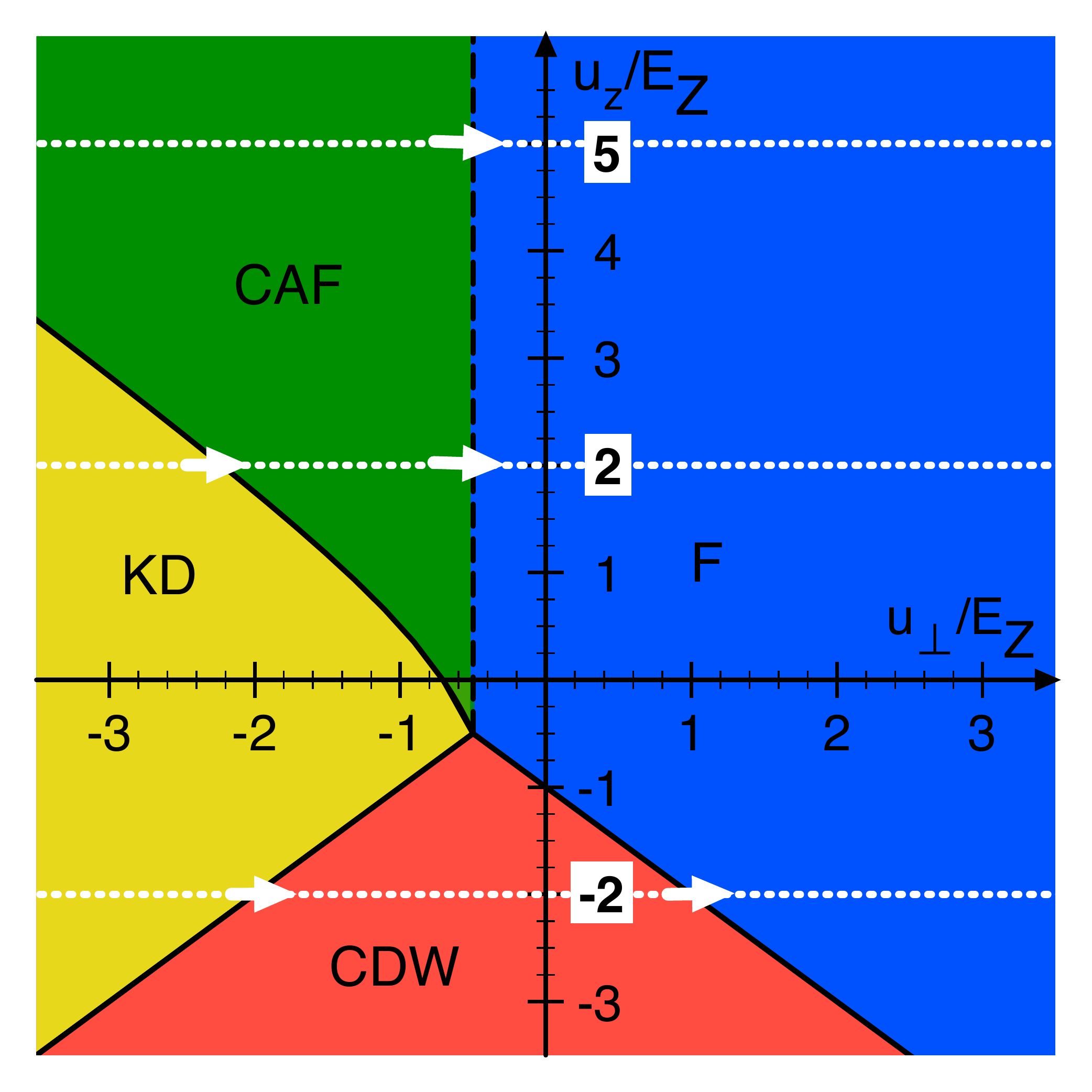}
  \caption{Bulk GS phase diagram as a function of the coupling 
energies $u_{\perp}$ and $u_z$ of the $\nu=0$ QH state for a Hamiltonian as in 
Eq.~(\ref{eqn:Htot}). 
Different colors distinguish between the different possible GS phases. The 
white, dotted lines indicate the parameters we choose throughout this paper to 
perform cuts through the phase diagram in order to explore the behavior of all 
possible GS phases when starting from the bulk and moving towards an edge of the 
graphene lattice. The phase diagram for the different GS phases in the bulk of graphene was first presented in Ref.~\onlinecite{kharitonov_phase_2012}.}
\label{fig:PhaseDiag_Khar}
\end{figure}

The total Hamiltonian we use is given by~:
\begin{equation}
\Hh=\Hh_{kin}+\Hh_{Z}+\Hh_{Coul}+\Hh_{Aniso},
\label{eqn:Htot}
\end{equation}
where we have a space dependent kinetic energy
induced by the presence of the boundary~:
\begin{equation}
\Hh_{kin}=   
-\sum_{i}E_{kin}(\mathbf{r}_i)\mathbf{\tau}^i_x,
\label{eqn:Hkin}
\end{equation}
{where the index $i$ labels the positions of the electron orbits.}
The hexagonal graphene lattice can be terminated in many different ways, yielding 
several possible edge structures. Every different atomic configuration leads to
different boundary conditions for the 
wavefunction\cite{akhmerov_boundary_2008}. Two extreme cases
 are the so-called \emph{zigzag} and \emph{armchair} 
edges~\cite{brey_electronic_2006}. A finite piece of  graphene 
 terminated by a zigzag edge 
and an armchair edge is shown in the 
inset of Fig.~\ref{fig:Ekin_vsX_wGrLatt}.
The kinetic energy and the corresponding eigenstates can be 
 obtained analytically~\cite{abanin_spin-filtered_2006, brey_edge_2006, mei_harmonic_1983, janssen_introduction_1994}. 
This is equivalent to turning the level index into a space dependent quantity $n(R)$ where $R$ 
relates to the distance to the edge $r$ as $r=R{\ell_B}/{\sqrt{2}}$. In 
Fig.~\ref{fig:Ekin_vsX_wGrLatt} we show the spatial shape of the kinetic energy 
$E_{kin}$ obtained by this procedure, as we will use it in the subsequent 
calculations.
We write the kinetic energy as a space-dependent potential proportional to 
$\tau_x$ (a comparable treatment can be found in Refs.~\onlinecite{kharitonov_edge_2012,murthy_collective_2014}). 
This corresponds to a 
perturbative treatment as it assumes an expansion of the 
perturbed edge states in terms of the unperturbed bulk basis states. It restricts our description to the case of "armchair-like" boundaries~: 
one can always infer the number of branches in the SP 
edge spectrum as being equal to the number of degenerate SP levels in the bulk and hence apply a perturbative expansion as 
implied by Eq.~(\ref{eqn:Hkin}). A derivation of such a Hamiltonian describing the kinetic potential of a graphene edge using 
arguments of perturbation theory can be found in Ref.~\onlinecite{kharitonov_edge_2012}. 
The edges terminated by a zigzag boundary condition support dispersionless surface states~\cite{brey_electronic_2006,brey_edge_2006}
that break the simple bulk/edge correspondence. They are beyond our simple treatment.
The form 
of the kinetic energy in Eq.~(\ref{eqn:Hkin}) is valid only in 
the regime $E_{kin}\ll\hbar\omega_{c}$, i.e., spatially not too close to 
the edge. As can be seen from Fig.~\ref{fig:Ekin_vsX_wGrLatt}, this condition is 
very well met if we restrict the subsequent discussion to the regime $R>3$.
Hence the restriction $R>3$ corresponds to a minimal distance 
$r_{min}\approx2.12\ell_B$, which at realistic experimental values  
corresponds to $r_{min}\approx120\text{a}_0$, where $\text{a}_0$ denotes the lattice constant 
of graphene. 
The Zeeman energy can be written as~:
\begin{equation}
\Hh_{Z}=   
-E_{Z}\sum_{i}\mathbf{\sigma}^i_z.
\label{eqn:HZ}
\end{equation}
In the case of graphene  the spacing between 
kinetic LL energy levels, $\Delta E_{kin}$, easily exceeds the Zeeman energy by two 
orders of magnitude.
The Coulomb interaction is given by~:
\begin{equation}
\Hh_{Coul} =  \frac{1}{2} \sum_{i\neq j}\frac{e^2}{\varepsilon}\frac{1}{|\mathbf{r}_i-\mathbf{r}_j|},
\label{eqn:HCoul}
\end{equation}
where $\varepsilon$ is an effective dielectric constant which depends upon the substrate\cite{santos_electric-field_2013}.
It has full SU(4) symmetry. At neutrality we have an example of quantum Hall ferromagnetism~\cite{MacDonald06}
with this large symmetry~: the ground state is thus highly degenerate and forms an irreducible
representation of SU(4). However this symmetry
is only approximate. In fact it is weakly broken by lattice-scale effects that
include short-range Coulomb interactions and electron-phonon couplings.
It is difficult to obtain precise estimates of these effects but their symmetry-breaking
properties can be encoded in the following Hamiltonian~:
\begin{equation}
\Hh_{Aniso} = \frac{1}{2} \sum_{i\neq j}\, \left[g_x\tau_x^i\tau_x^j+g_y\tau_y^i\tau_y^j+g_z\tau_z^i\tau_z^j\right]\,\,
\delta^2 (\mathbf{r}_i-\mathbf{r}_j),
\label{eqn:HAniso}
\end{equation}
with $\delta$ denoting the Dirac delta function. This Hamiltonian $\Hh_{Aniso}$ 
has been proposed by Aleiner \textit{et al.}~\cite{aleiner_spontaneous_2007}
 and its effects have been analyzed at the mean-field level by Kharitonov\cite{kharitonov_phase_2012}. 
Its symmetry properties and phase diagram have been studied by exact diagonalization~\cite{Wu14}.
It is parametrized by the coupling constants $g_{x,y,z}$ whose values are not known with precision.
It is likely that the ratio of the energy scales between Coulomb interaction and these anisotropies
is of the order of $10^2$. It is thus best to explore the complete phase diagram taking these
parameters as unknowns.
For monolayer and bilayer graphene at neutrality, there is a rich phase diagram 
as a function of these couplings~\cite{kharitonov_phase_2012, kharitonov_canted_2012}. 
The fractional quantum Hall states are also sensitive to these effects~\cite{sodemann_broken_2014}.

We now perform a HF study of this Hamiltonian by including the edge potential.
We note that in this approach we neglect all possible spatial dependence of the 
coupling constants, which is justified as long as we analyze a 
spatial domain not too close to the edge.

\section{Hartree-Fock treatment}
\label{section:HF_treat}
The neutral $\nu=0$ state 
corresponds to the half-filled case where two of the four available states per 
orbital are occupied. We look for the ground state within the family of 
 Slater determinants of the form~:
\begin{equation}
|G\rangle=\prod_{p}
\left(\sum_{\mu,\nu}\, 
g_{\mu\nu}\,c_{\mu}^{\dagger}(p)c_{\nu}^{\dagger}(p)
\right)\,|0\rangle,
\label{eqn:GroundState}
\end{equation}
where $p$ denotes the Landau-gauge momentum component along the edge which labels the electron orbitals. The vacuum $|0\rangle$ consists of the 
completely occupied set of states 
for all $n<0$ and completely empty states for all $n>0$. In 
Eq.~(\ref{eqn:GroundState})  ${{g}}$ is a 4 by 4 antisymmetric matrix, i.e., $g_{\mu\nu}=-g_{\nu\mu}$, in order to describe a valid 
fermionic state and $\Tr[gg^{\dagger}]=2$ to ensure normalization of the 
two-particle state. 
We minimize the energy of the Slater determinant by varying $g$.
To capture the effect of the edge potential we take the $g$ matrix to be momentum dependent,
i.e., $g\equiv g(p)$ in Eq.~(\ref{eqn:GroundState}). Due to the duality 
between the longitudinal momentum $p$ and the transverse coordinate $r_p=p\ell^2_B$ this is equivalent to a space dependent description of the problem.
The most general antisymmetric matrix ${{g}}$ has 12 real 
 parameters. By exploiting the symmetry properties of 
 the state $|G\rangle$ and the Hamiltonian, one can reduce the 
number of free parameters. We use the same strategy as in 
Ref.~\onlinecite{ezawa_ground-state_2005} 
where an equivalent problem was studied in the context of electronic bilayer systems.
It is convenient to rewrite the problem in terms of the following simple expectation values~:
\begin{subequations}
\begin{align}
S_{\alpha}&=\frac{1}{2}\langle G| 
\cc^{
\dagger}(p)\mathbf{\sigma}_{\alpha}\,\cc(p)
|G\rangle=\frac{1}{2}\Tr[\sigma_{\alpha}gg^{\dagger}], \label{eqn:Sa}\\
T_{\alpha}&=\frac{1}{2}\langle G| 
\cc^{
\dagger}(p)\mathbf{\tau}_{\alpha}\cc(p)
|G\rangle=\frac{1}{2}\Tr[\tau_{\alpha}gg^{\dagger}], \label{eqn:Na}\\
R_{\alpha\beta}&=\frac{1}{2}\langle G| 
\cc^{\dagger}(p) \mathbf{\sigma}_{\alpha}\mathbf{\tau}_{\beta}\cc(p)
|G\rangle=\frac{1}{2}\Tr[\sigma_{\alpha}\tau_{\beta}gg^{
\dagger}].
\end{align}
\label{eqn:SNR}
\end{subequations}
The expressions of Eqs.~(\ref{eqn:Sa}) and (\ref{eqn:Na}) yield the components of 
the total spin $S_{\alpha}$ and isospin $T_{\alpha}$ per orbital $p$.
The energy contribution from the symmetry breaking interaction 
Hamiltonian of Eq.~(\ref{eqn:HAniso}) is now given by~:
\begin{equation}
\langle G| \Hh_{Aniso} |G\rangle= \frac{1}{2}\sum_{\alpha}u_{\alpha}\Big( 
\Tr[\tau_{\alpha}gg^{\dagger}]^2 
-\Tr[\tau_{\alpha}gg^{\dagger}\tau_{\alpha}gg^{\dagger}]  \Big),
\label{eqn:Ebreak}
\end{equation}
where the anisotropy energies 
$u_{\alpha}$ are given by
$u_{\alpha}=\frac{g_{\alpha}}{2\pi\ell_B^{2}}$. 
Isotropy of the interaction potential in the $x$-$y$-plane 
implies that $u_x=u_y=:u_{\perp}$.
Using Eqs.~(\ref{eqn:SNR}) and (\ref{eqn:Ebreak}), we obtain
the following expression for the functional of the total energy $E_{tot}=\langle 
G|\Hh|G\rangle$~:
\begin{equation}
E_{tot}=-2E_{kin}\, T_x-2E_Z \, 
S_z+\sum_{\alpha}u_{\alpha}\Big(T_{\alpha}^2-\sum_{i}R^2_{i\alpha}-\mathbf{S}^2  
 \Big).
\label{eqn:Etot}
\end{equation}

The $12-2=10$ free parameters of the problem (dropping the 
overall phase and normalization constant) are now encoded in the 6 components of 
the total spin $\mathbf{S}$ and the total isospin $\mathbf{T}$, together with 4 
out of 9 components of $R_{\alpha\beta}$ which can be chosen independently. The 
invariance of $E_{tot}$ in Eq.~(\ref{eqn:Etot}) under rotations of 
$\mathbf{S}$ in spin space and rotations of $\mathbf{T}$ around the $z$-axis in 
isospin space allows us to choose $S_y=S_x=T_y=0$ with no 
loss of generality, yielding seven variables to be determined. The dimension 
of parameter space can be further reduced by careful consideration of all the 
symmetries of the problem. As demonstrated by Ezawa 
\textit{et al.}~\cite{ezawa_ground-state_2005} in a situation of an equivalent 
symmetry class, reduction is possible to a total number of three free 
parameters. For the present system, this leads us to a minimization problem for 
the total 
energy $E_{tot}$ with respect to the variational 
parameters $-1\leq\alpha\leq1, -1\leq\beta\leq 1$, and $\chi\in\mathbb{R}$, 
which are related to observables of Eq.~(\ref{eqn:SNR}) by~:
\begin{equation}
S_z=\frac{1}{\sqrt{1+\chi^2}}\,\sqrt{1-\alpha^2},\quad
T_x=\frac{\chi}{\sqrt{1+\chi^2}}\,\alpha\sqrt{1-\beta^2},\quad
T_z=\frac{\chi}{\sqrt{1+\chi^2}}\,\alpha\beta,
\label{eqn:SzNzNz_abz}
\end{equation}
and 
\begin{equation}
\sum_{i}R^2_{ix}=\frac{T_z^2}{\chi^2},\quad
\sum_{i}R^2_{iy}=\chi^2 \mathbf{S}^2,\quad
\sum_{i}R^2_{iz}=\frac{T_x^2}{\chi^2},
\label{eqn:R_abz}
\end{equation}
where the index $i$ runs over the spatial components $\{x,y,z\}$.
The density matrix $\rho_g=gg^{\dagger}$ is connected to these quantities as (summation convention implied)~:
\begin{equation}
\rho_g=\frac{1}{2}\mathbb{1}+\frac{1}{2}\Big( \sigma_i\, S_i +\tau_i\, T_i 
+\sigma_i\tau_j\, R_{ij}    \Big).
\label{eqn:rho_SNR}
\end{equation} 

\section{Ground state properties}
\label{section:PropGS}

\subsection{Evolution of the Spin-Isospin Texture close to the Edge}
The bulk GS  of graphene within the model we use has several different phases 
depending on the anisotropy energies $u_{\perp}$ and $u_{z}$ compared to the 
Zeeman energy $E_Z$\cite{kharitonov_phase_2012,Wu14}. 
The anisotropies $u_{\perp}, u_{z}$ and the Zeeman term $E_Z$
select some subset of the manifold of SU(4) ferromagnetic ground states.
These phases 
have distinct spin and isospin configurations, i.e., by 
different spin textures.
The mean-field GS diagram is shown in Fig.~\ref{fig:PhaseDiag_Khar}.
It is correct beyond mean-field as shown by exact diagonalization techniques~\cite{Wu14}.
Notably all these bulk phases  do not involve spin/valley entanglement. 
We now generalize the description of these
phases by treating the effect of a boundary of the lattice. 
We proceed as follows~: we minimize the energy 
$E_{tot}(\alpha,\beta,\chi)$ of Eq.~(\ref{eqn:Etot}), including a space dependent 
edge potential of the shape shown in Fig.~\ref{fig:Ekin_vsX_wGrLatt} by varying 
the parameters $\alpha,\beta, \chi$. Then from the knowledge of
 the parameters $\alpha(R),\beta(R), \chi(R)$, we 
can compute the values of the observables $S_z(R), T_x(R), T_z(R)$ of the GS 
$|G\rangle$ via Eq.~(\ref{eqn:SzNzNz_abz}). It is also possible to construct the entire
density matrix $\rho_g$ characterizing the GS via Eq.~(\ref{eqn:rho_SNR}).

In order to obtain a full picture capturing the edge behavior of all possible 
bulk phases shown in Fig.~\ref{fig:PhaseDiag_Khar}, our
choice of system parameters is guided by the following idea~: for fixed Zeeman 
energy $E_Z$, we vary the coupling energies $u_{\perp}$ and $u_z$ because
this can be realized experimentally by tilting the magnetic field. 
We choose
three values of the perpendicular coupling energy $u_z$~: $u_z=5E_Z, 
\; u_z=2E_Z$ and $u_z=-2E_Z$, and we vary the perpendicular coupling  
in the range $-3E_Z\le u_{\perp} \le 3E_Z$. This leads to horizontal 
cuts through the $\nu=0$ GS phase diagram in the $u_{\perp},u_z$-plane,
 shown by  white dotted lines in 
the phase diagram of Fig.~\ref{fig:Ekin_vsX_wGrLatt}. 
For $ u_z=5E_Z$ and $u_z=2E_Z$ by varying $u_\perp$ we meet the KD, CAF and F phases~:

\begin{tikzpicture}
 \draw[white] (0,0) -- (3,0); \draw[gray,thick] (3,0)node[left]{\textcolor{black}{$u_{\perp}=-\infty$}} 
node[above]{\hskip20pt KD} --node[below]{}(5,0); \draw[teal,thick] (5,0)--node[below]{\textcolor{black} {$u_{\perp} > 
-\frac{1}{2}\big(u_z+\sqrt{2E^2_Z+u_z^2}\big)$} } node[above]{CAF} (10,0) ; 
\draw[blue,thick,->] (10,0)--node[align=right,below]{\textcolor{black}{$u_{\perp}>-\frac{E_Z}{2}$}} 
node[above]{F} (14,0)node[right]{\textcolor{black}{$u_{\perp}$}} ;
\draw[thick](5,-0.2)--(5,0.2);\draw[thick](10,-0.2)--(10,0.2);
\end{tikzpicture}

For $u_z=-2E_Z$ and varying again $u_\perp$ we find the KD, CDW and F phases~:

\begin{tikzpicture}
 \draw[white] (0,0) -- (3,0); \draw[gray,thick] (3,0)node[left]{\textcolor{black}{$u_{\perp}=-\infty$}} 
node[above]{\hskip20pt KD} --node[below]{}(4,0); \draw[red,thick] (4,0)--node[below]{\textcolor{black} 
{$u_{\perp} > u_z$} } node[above]{CDW} (11,0) ; \draw[blue,thick,->] (11,0)--node[align=right,below]
{\textcolor{black}{$u_{\perp}>-(E_Z+u_z)$}} node[above]{F} (14,0)node[right]{\textcolor{black}{$u_{\perp}$}} ;
\draw[thick](4,-0.2)--(4,0.2);\draw[thick](11,-0.2)--(11,0.2);
\end{tikzpicture}

The corresponding bulk phase transitions  are indicated by white 
arrows in the phase diagram in Fig.~\ref{fig:Ekin_vsX_wGrLatt}. 
\begin{figure}[t]
\centering
\includegraphics[width=0.6\textwidth]{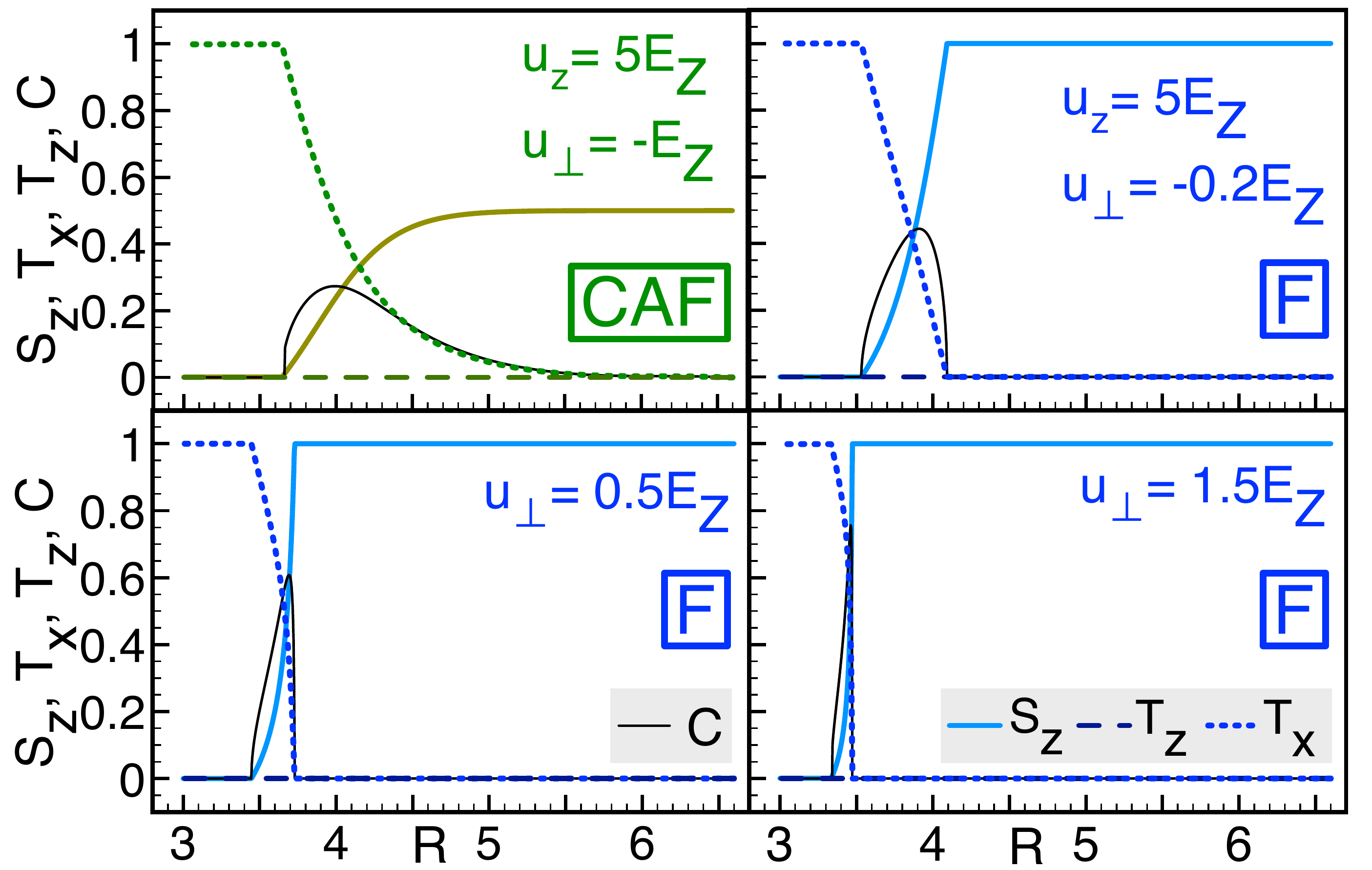}
\caption{Evolution of the spin and isospin 
 as well as the concurrence $C$ as functions of 
$R=\sqrt{2}r/\ell_B$, with $r$ the distance from the edge.
We fix $u_z=5\,E_Z$ and vary $u_{\perp}$. 
(Colored lines: $S_z$, solid; $T_{z}$, dashed; $T_{x}$, dotted. 
Black, thin, solid line: Concurrence $C$).
There is a transition between the bulk phase on the right-hand side
 [CAF for $u_{\perp}=-E_Z$ in the upper left panel 
(green colors) and F for all other choices of $u_{\perp}$ shown (blue colors)] 
to a KD phase at the edge. In an intermediate regime
 spin and isospin, are canted with $0<S_z<1$ and 
$0<T_x<0$ at the same time and nonzero values of the concurrence.
With growing $u_{\perp}$, this domain wall grows narrower in space and moves 
closer to the boundary. }
 \label{fig:Obs_AllPhases}
\end{figure}

\begin{figure}[t]
\centering
\includegraphics[width=0.6\textwidth]{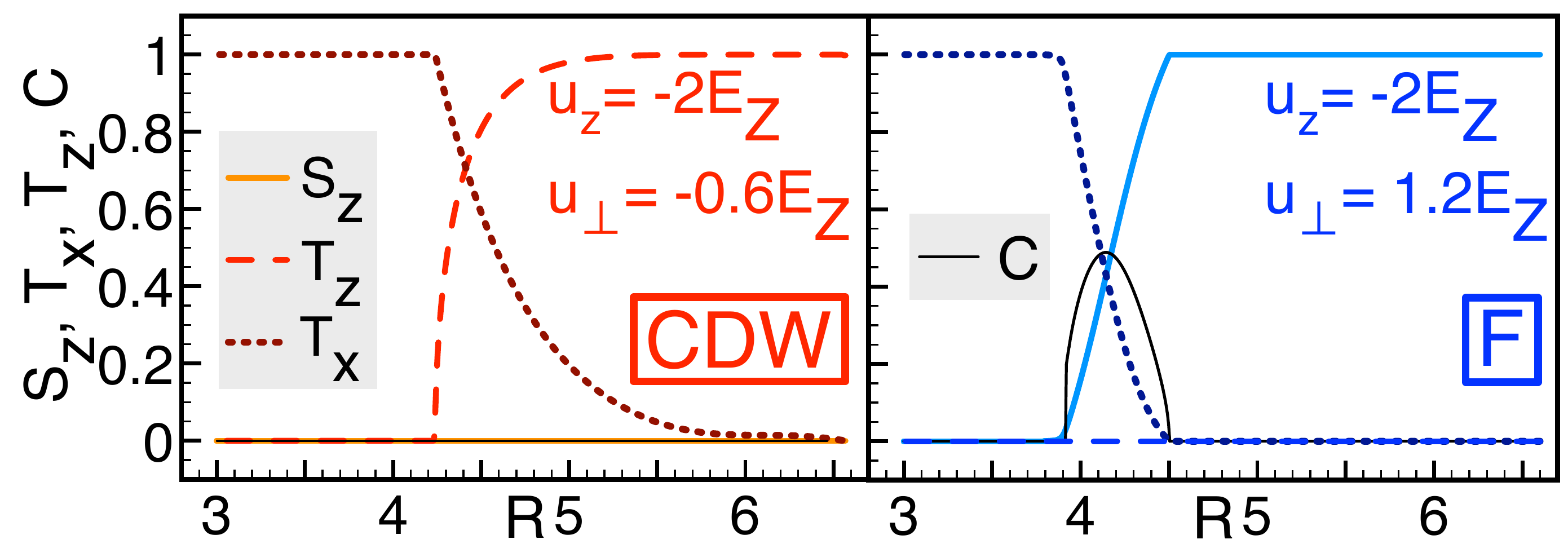}
\caption{Same as Fig.~\ref{fig:Obs_AllPhases} for transverse coupling energy 
$u_z=-2 E_Z$ and different perpendicular coupling energies $u_{\perp}$, such 
that the phases of the lower half plane of the GS phase diagram are established 
in the bulk. Left panel: $u_{\perp}=0.6$, leading to a CDW in the bulk (red 
colors). Right panel: $u_{\perp} =1.2E_Z$, at which the bulk is in a F phase 
(blue colors).}
\label{fig:Obs_AllPhases_LHP}
\end{figure}

We first investigate the influence of the edge potential on the spin and isospin observables
$\mathbf{S}$ and $\mathbf{T}$. More precisely, we discuss the spatial evolution of the components $S_z(R), 
T_x(R), T_z(R)$ for different choices of the anisotropy energies $u_{\perp}$ and 
$u_{z}$ compared to the Zeeman energy $E_Z$.
Figure \ref{fig:Obs_AllPhases} shows the results for $u_z=5E_Z$, 
corresponding to a cut through the upper half plane (we plot obsevables with colored lines). The four 
different panels depict the situation for different values of  
 $u_{\perp}$~: for $u_{\perp}=-E_Z$, the bulk system 
is in the CAF phase with canting angle $\cos{\theta}=S_z=1/2$ (upper left panel), 
whereas for $u_{\perp}=-0.2E_Z$, $u_{\perp}=0.5E_Z$, and $u_{\perp}=1.5E_Z$ the 
bulk system establishes a F phase in which the spins are fully polarized. 
The colored curves shown in Fig.~\ref{fig:Obs_AllPhases_LHP} correspond to the obervables for a
cut through the lower half plane of the phase diagram at $u_{z}=-2E_Z$. Here, 
the perpendicular couplings are chosen so that the left panel at 
$u_{\perp}=-0.6E_Z$ corresponds to a CDW phase 
whereas the right panel at $u_{\perp}=1.2E_Z$ again corresponds to a F bulk 
phase, as predicted by the GS phase diagram of Fig.~\ref{fig:PhaseDiag_Khar}. Curves for 
values of the anisotropy energies favoring a KD phase in the bulk are not 
shown since in this case the system does not undergo any transition whatsoever
but remains in the bulk KD phase all the way to the edge.

In general, one can distinguish three different regimes for the 
behavior of the observables as a function of the distance $r=\frac{\ell_B}{\sqrt{2}}R$ to 
the edge.
For sufficiently large values of $R$, i.e., deep enough in the bulk, we 
recover the results of mean-field theory~\cite{kharitonov_phase_2012}. 
Close enough to the edge, the 
system is driven into a KD phase with $N_x=1$ and $S_z=N_z=0$, independently of 
the bulk phase it adopts. This behavior is due to the edge potential 
 in the kinetic energy Hamiltonian $\Hh_{kin}$ in 
Eq.~(\ref{eqn:Hkin})~: this term is proportional to $\tau_x$, so it acts as a 
Zeeman effect in isospin space, polarizing the isospin along the $x$-direction 
 as soon as $E_{kin}(R)$ is large enough.
 This behavior is also consistent with previous works~\cite{fertig_luttinger_2006, murthy_collective_2014}. 
In an intermediate regime we find a finite interval in space in 
which $S_z\neq1,\, T_x\neq 1$ and $T_z\neq0,\, N_x\neq 0$, i.e., 
the spin and the isospin are canted simultaneously with respect to 
their bulk values. There is thus a domain wall at a 
small finite distance from the edge. For 
the CAF configuration, this domain wall connects smoothly to the bulk configuration. For a system 
in a F phase in the bulk however, the change in spin and isospin is abrupt
and the domain wall is
narrower with increasing $u_{\perp}$. Hence, for larger values of $u_{\perp}$, the F 
phase of the bulk proves to be more resistant against the increasing influence 
of the edge.

From our analysis and the results shown in Figs.~\ref{fig:Obs_AllPhases} and 
\ref{fig:Obs_AllPhases_LHP} we therefore draw the following conclusions: The 
phases in the bulk of a finite sample of graphene do not remain unaffected 
close enough to the edges. Indeed the  effective edge potential 
causes the bulk state to undergo a transition in which 
 the polarizations of spin and isospin change simultaneously. 
 Sufficiently close to the edge,
the GS is driven into a KD phase independently of the nature of the bulk phase.

\subsection{Spin-Valley Entanglement of Edge States}
\label{sssection:SVent}

\begin{figure}[t]
  \centering
\includegraphics[width=0.6\textwidth]{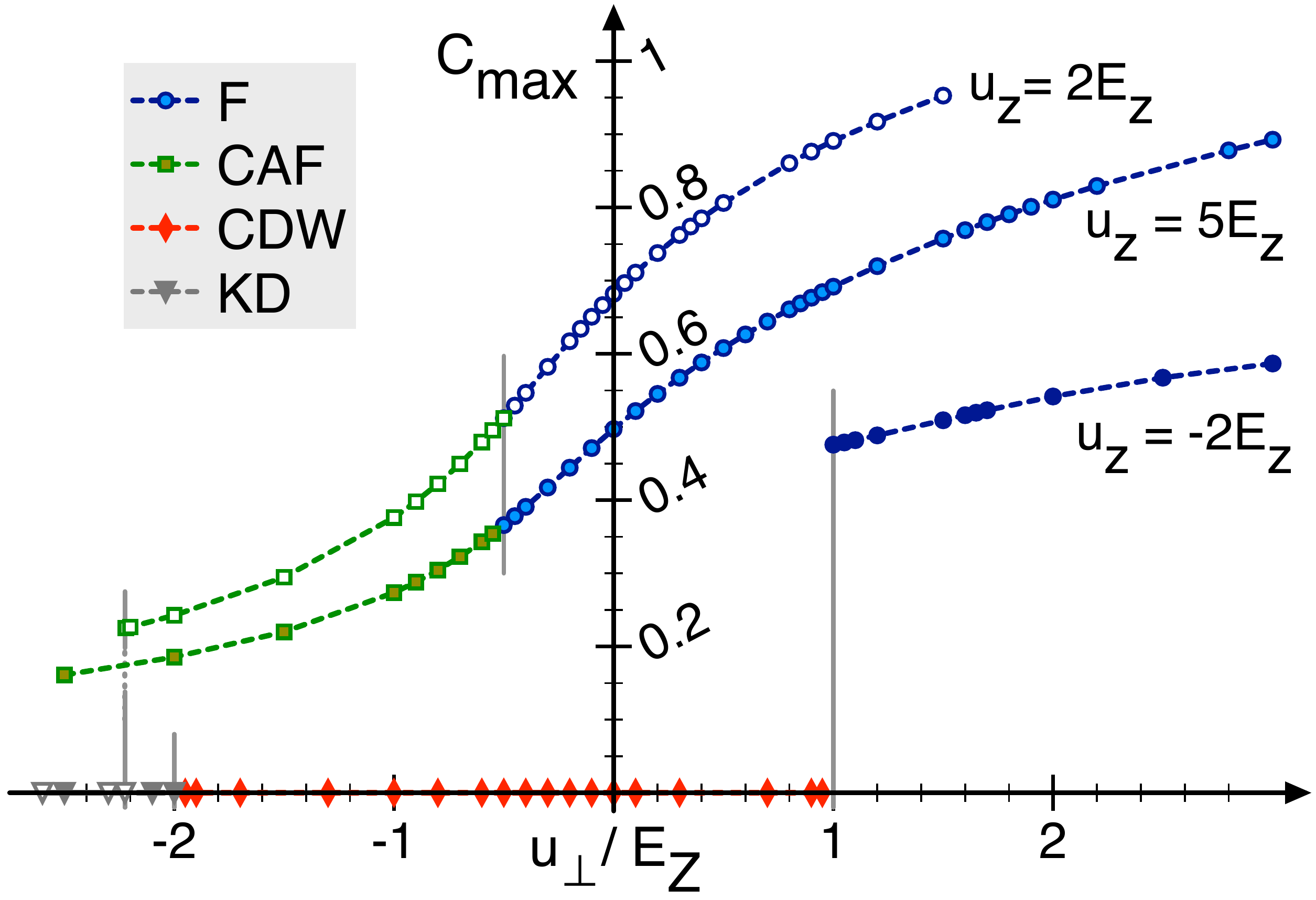}
  \caption{Maximum concurrence $C_{max}$  in the different regimes 
 from the bulk to the edge. For three different values of 
the coupling energy $u_{z}$, we vary $u_{\perp}$. 
Different symbols represent different phases in 
the bulk at 
the respective value of $u_{\perp}$: F (blue circles), CAF (green squares), CDW 
(red diamonds), and KD (gray triangles). Empty, shaded, or filled symbols 
distinguish between the cuts at different $u_{z}$. The dashed 
lines connect the data points as a guide to the eye. Vertical, gray lines mark 
the values of $\{u_z,u_{\perp}\}$, at which, in the bulk, transitions between the 
respective phases occur.
We observe that non-zero values of $C$ happen only if the system in the bulk 
is in a CAF or in a F phase. In this case, the CAF/F transition is smooth. If the bulk 
is CDW or KD,the concurrence remains equal to zero 
all the way from the bulk to the edge. This leads to a discontinuous jump in the 
curve of $C_{max}$ at the point of the CDW/F transition.}
    \label{fig:MaxConc_vsu}
\end{figure}

The parametrization of $\mathbf{S}, \mathbf{T},$ and 
${{\text{R}_{\alpha\beta}}}$ in terms of the three parameters $\alpha,\beta,$ 
and $\chi$  in Eqs.~(\ref{eqn:SzNzNz_abz}) and (\ref{eqn:R_abz}) 
allows for complete reconstruction of the density matrix $\rho_g$ via 
Eq.~(\ref{eqn:rho_SNR}) using the values of the parameters obtained by minimizing 
$E_{tot}$ from Eq.~(\ref{eqn:Etot}). Therefore, we have a full 
description of the GS and its spatial evolution from the bulk towards the 
edge. 
In this paragraph we  investigate the \emph{entanglement} between spin 
and isospin degrees of freedom in the system. For the infinite bulk case, product states of 
the form $|s\rangle\otimes |n\rangle$, where $|s\rangle$ denotes the (single 
particle) spin state and $|n\rangle$ the (single particle) isospin state, have 
been used as an ansatz to minimize the GS energy\cite{kharitonov_phase_2012, 
kharitonov_canted_2012, kharitonov_edge_2012}. Existing studies of edge states 
using a variational trial wave function approach have
suggested~\cite{murthy_collective_2014}, however, that for a non-zero edge 
potential, spin and isospin might not remain independent, separable observables, 
but become \emph{entangled}.
In order to quantify the amount of entanglement in the bipartite two-level 
system $H=H_{spin}\otimes H_{valley}$, we calculate the concurrence $C$ 
according to the definition\cite{mintert_measures_2005}~:
\begin{eqnarray}
C=\text{max}(\lambda_1-\lambda_2-\lambda_3-\lambda_4,0),
\label{eqn:Def_C}
\end{eqnarray}  
where the $\lambda_i$ are the eigenvalues of the matrix
\begin{equation}
\mathfrak{R}=\sqrt{\rho_g}(\sigma_y\otimes\sigma_y)\rho_g^*(\sigma_y\otimes\sigma_y)\sqrt{\rho_g},
\label{eqn:Def_RforC}
\end{equation}
in decreasing order $\lambda^2_i\geq\lambda_{i+1}^2\;\forall i$. In Eq.~(\ref{eqn:Def_RforC}), 
$\sigma_y$ denotes the $2\times2$ Pauli matrix. The quantity 
$C$ ranges from 0 to 1 with $C=0$ meaning no entanglement and $C=1$ for 
maximally entangled states.

In order to study the entanglement of all phases, we 
perform the same cuts through the phase diagram as in the previous paragraph, 
fixing the transverse coupling at $u_z=5E_Z$ and $u_z=2E_Z$ to 
investigate the upper half plane and at $u_z=-2E_Z$ to study the lower half 
plane and we vary the perpendicular coupling  $u_{\perp}$ at these fixed values.
Examples of the spatial behavior of the concurrence $C(R)$ as a function of 
the distance from the edge is depicted by the black solid lines in 
Figs.~\ref{fig:Obs_AllPhases} and \ref{fig:Obs_AllPhases_LHP}. The curves reveal 
several characteristics of the behavior of the concurrence. 
It goes to  zero deep enough in the bulk for all values of 
the anisotropies $\lim_{R\to\infty}C(R)=0\; \forall\; u_z,u_{\perp}$. 
Close enough to the edge, the concurrence is also equal to zero for all possible
bulk phases, as can be seen in Figs.~\ref{fig:Obs_AllPhases} and 
\ref{fig:Obs_AllPhases_LHP} for $C(R\approx3)\equiv0\; \forall \; u_z,u_{\perp}$. 
In an intermediate regime for
which the system is in a F or CAF phase in the bulk, we find that the concurrence  
develops a sharp peak. This
peak appears precisely within the domain wall separating 
its bulk phase to the KD phase near the edge. The peak is 
sharper and higher with rising $u_{\perp}$, as the domain wall becomes more and 
more narrow in space.
Another behavior is observed when the bulk is CDW
 (left panel of Fig.~\ref{fig:Obs_AllPhases_LHP}). 
 Here, the concurrence remains zero 
independently of the distance from the edge~: $C(R)\equiv0\;\forall R$.

 Our findings are summarized in Fig.~\ref{fig:MaxConc_vsu}, where
we plot the maximum 
concurrence $C_{max}$ as a function of $u_\perp$. The resulting curves 
characterize the behavior of the spin-valley entanglement of the edge states.
Nonzero values of the concurrence are found only for anisotropies 
favoring a CAF phase (green squares) or a F phase (blue circles) in the bulk. 
In these regimes, the maximum concurrence $C_{max}$ is a 
monotonically rising  function of the perpendicular coupling 
$u_{\perp}$, with no discontinuity at $u_{\perp}=-E_Z/2$, which would correspond to the
CAF/F transition in the bulk.  Discontinuous jumps
appear at values of $u_{\perp}$ corresponding to the transitions KD/CAF or CDW/F.
Combining the information from Figs.~\ref{fig:Obs_AllPhases}, 
\ref{fig:Obs_AllPhases_LHP} and Fig.~\ref{fig:MaxConc_vsu}, 
 we draw the following conclusions~: unlike the states in an 
infinite system, the GS in the presence of a  boundary may exhibit 
nonzero spin-valley entanglement. As demonstrated in 
Figs.~\ref{fig:Obs_AllPhases} and \ref{fig:Obs_AllPhases_LHP}, the concurrence 
is exactly zero in all configurations where either the spin or 
the isospin is strictly zero. Nonzero values of the 
concurrence appear for configurations in which both spin and  isospin 
are canted \textit{simultaneously}. 

Compared to the bulk case, we find that  the lattice 
boundary  gives rise to novel ground state phases close to the edge, 
with simultaneous canting of spin and isospin, $0<\mathbf{S}<1, \; 
0<\mathbf{T}<1 $, and nonzero spin-valley entanglement. These phases cannot
 be described using trial wave functions in the form of separable product 
states~\cite{kharitonov_phase_2012, kharitonov_edge_2012}. 
 
\section{Mean Field Spectrum of Excited States}
\label{ssection:MFSpectrum}

\subsection{Single Particle Energy Levels}
\label{sssection:SPLevels}

\begin{figure}[t]
  \centering
\includegraphics[width=0.6\textwidth]{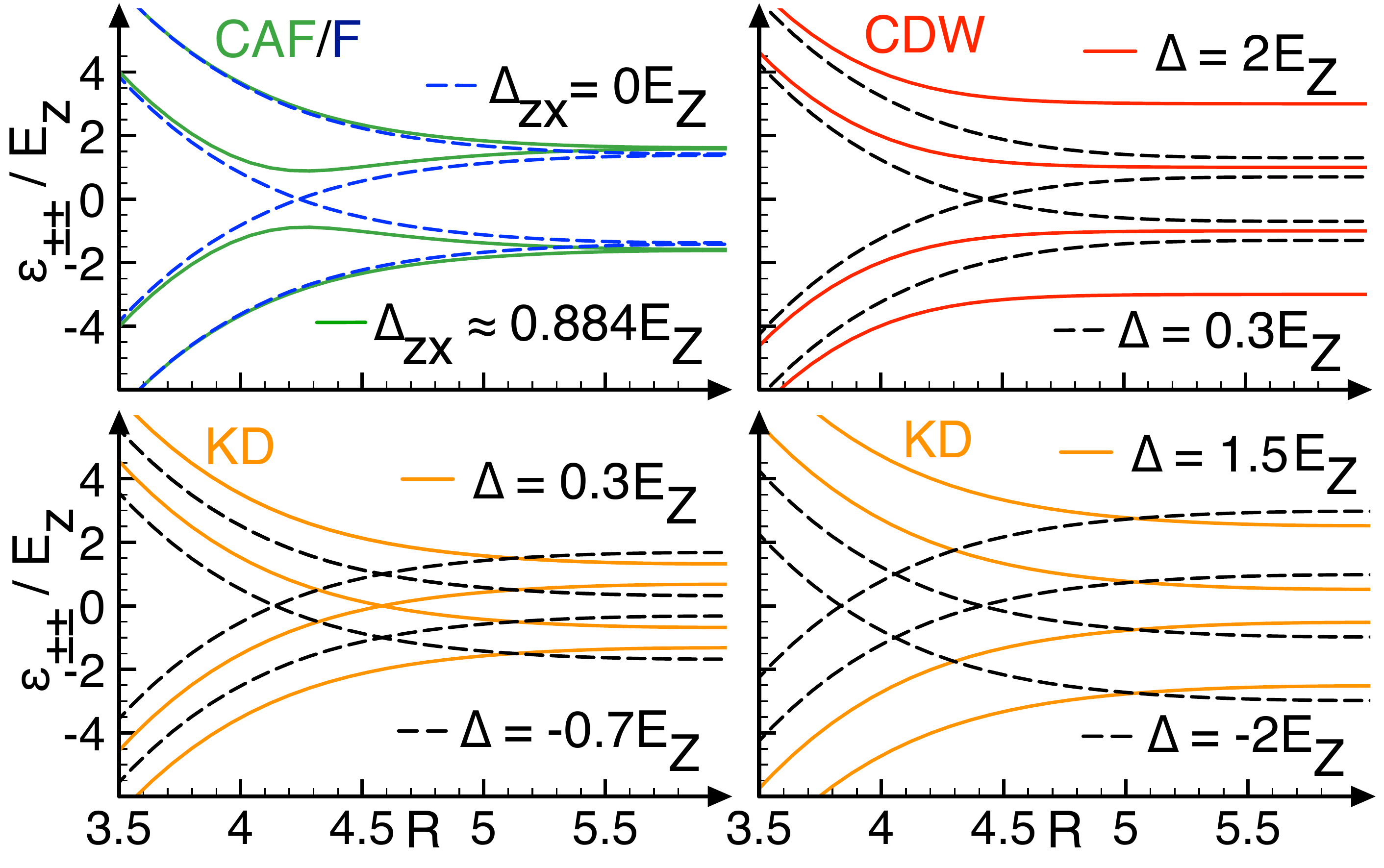}
  \caption{SP energy levels $\varepsilon_{\pm\pm}$ for the different phase 
regimes as they are obtained from the analytical formulas in Table \ref{tab:AnalyticExpr}. 
 Upper left panel: CAF (solid, green lines), F (blue, dashed lines). Upper right 
panel: CDW for $|\Delta|>E_Z$ (solid, red lines) and $|\Delta|<E_Z$ (dashed, 
black lines). Lower left panel: KD for $|\Delta|<E_Z$ with $\Delta>0$ (solid, 
orange lines) or $\Delta<0$ (dashed, black lines). Lower right panel: KD for 
$|\Delta|>E_Z$ with $\Delta>0$ (solid, orange lines) or $\Delta<0$ (dashed, 
black lines).
Depending on the sign and the magnitude of $\Delta$, the different cases for the 
SP spectra  differ in the number of level crossings, even within one and the 
same phase.}
    \label{fig:SPL_anltc}
\end{figure}

We now
diagonalize the single-particle HF Hamiltonian.  The spectrum of the excited states 
is of particular interest, since the conduction properties of real graphene samples
are governed by the edge modes in the QH regime.
Recent conductance experiments have 
shown~\cite{young_tunable_2014} that upon tilting the applied magnetic field  
there is a transition from an insulating regime 
 to a phase where presumably edge states carry a nonzero current. Tilting 
the magnetic field corresponds to  varying the parameter 
$u_{\perp}/E_Z$ in the system. The 
experimental observations therefore suggest that the gap to excited states in the edge 
spectrum varies as a function of $u_{\perp}/E_Z$ and closes, eventually, 
giving rise to a metal-insulator transition.
We write the one-body HF Hamiltonian 
$\h^{HF}$
corresponding to the full Hamiltonian $\Hh$ of Eq.~(\ref{eqn:Htot}). It consists of four terms~:
\begin{equation}
\h_{\mu\nu}^{HF}(p)=-E_{kin}(p)[\tau_x]_{\mu\nu}- E_Z 
[\sigma_z]_{\mu\nu}+\;^C\Delta_{\mu\nu}+\;^A\Delta_{\mu\nu},
\label{eqn:hHF}
\end{equation}
which we obtain via standard HF decoupling. For the mean-field potential from the 
Coulomb interaction Hamiltonian of Eq.~(\ref{eqn:HCoul}) we find
\begin{equation}
^C\Delta_{\mu\nu}=-u_0 [gg^{\dagger}]_{\mu\nu},
\label{eqn:Coul_HF}
\end{equation}
where $u_0$ 
describes the exchange term of the Coulomb interaction Hamiltonian of 
Eq.~(\ref{eqn:HCoul}). This formula is valid provided we neglect the spatial dependence
of $g$. It means that we do not capture the spin texture effects of the exchange.
For completeness, in 
the following analytical calculations and expressions, the Coulomb contribution 
of Eq.~(\ref{eqn:Coul_HF}) will be written explicitly. 
The mean-field potential due to the interactions breaking SU(4) symmetry is given by~:
\begin{equation}
^A\Delta_{\mu\nu}= \sum_{\alpha} u_{\alpha} \Big( [\tau_{\alpha}]_{\mu\nu} \,  
\Tr[gg^{\dagger}\tau_{\alpha}] -[ \tau_{\alpha} gg^{\dagger} \tau_\alpha 
]_{\mu\nu} \Big).
\label{eqn:Break_HF}
\end{equation}
%
%
Within HF mean-field theory, diagonalizing $\h^{HF}$ provides access to SP energies 
$\varepsilon_i$, satisfying $\h^{HF}|i\rangle=\varepsilon_i |i\rangle$, where the state labeled by $i$ stands for the $i$th SP HF eigenstate. In the 
following, we  assume the eigenvalues to be ordered  
$\varepsilon_1\le\varepsilon_2\le\varepsilon_3\le\varepsilon_4$.
In earlier work~\cite{kharitonov_edge_2012}, this Hamiltonian 
 has been studied with the assumption of constant order parameter to the edge. 
However, the explicit effective valley field due to the edge certainly invalidates this simple assumption.
It should be noted also that in the intermediate regime between the bulk state and the edge regime
the HF ground state is no longer a simple tensor product state even within the HF approximation
and there is some nontrivial spin/valley entanglement.
To facilitate subsequent 
discussion we summarize the results of Ref.~\onlinecite{kharitonov_edge_2012} 
in Table \ref{tab:AnalyticExpr}. The analytical expressions for the mean-field levels were presented in 
Ref.~\onlinecite{kharitonov_edge_2012} for the  F/CAF cases. A straightforward 
calculation allows direct extension to the KD and the CDW configuration. Examples for 
the SP energy levels $\varepsilon_{\pm\pm}$ obtained within this approximation for the different 
phases are plotted in Fig.~\ref{fig:SPL_anltc} for various choices of the couplings.


\begin{table}[t]
\begin{tabular}{ l }
\hline\hline
CAF/F phase: \\
\hline\\
$^A\Delta=\;^A\Delta_{0z}\mathbb{1}\otimes\sigma_z+\;^A\Delta_{zx}\sigma_z\otimes\sigma_x$\\
with $^A\Delta_{0z}=-\frac{1}{2}(u_0+u_z+2u_{\perp})\cos{\theta}$ and $^A\Delta_{zx}
=-\frac{1}{2}(u_0+u_z-2u_{\perp})\sin{\theta}$,\\ \\
$\varepsilon_{\pm\pm}=\pm 
 \sqrt{\big[E_{kin}(p)\pm\big(E_Z-\;^A\Delta_{0z}\big)\,\big]^2+\big(\,
 ^A\Delta_{zx}\big)^2}$,\\ \\
 $\Delta\varepsilon_{edge}=2 \;|\,^A\Delta_{zx}|, 
$
$ \Delta\varepsilon^{CAF}_{bulk}=u_0+u_z-2u_{\perp},\label{eqn:BulkGap_CAF}$
$ \Delta\varepsilon^{F}_{bulk}=2|E_Z-\,^A\Delta_{0z}|$.\\ \\
\hline\hline
CDW/KD phase:\\ \hline\\
$ ^A\Delta =\;^A\Delta _{x0}\sigma_x\otimes\mathbb{1}+\;^A\Delta _{y0}\sigma_y\otimes\mathbb{1}+\;^A\Delta_{z0}
 \sigma_z\otimes\mathbb{1}$\\
 with $^A\Delta _{x,y0}=-\frac{1}{2}(u_0t_{x,y}-u_zt_{x,y}-4u_{\perp}t_{x,y})$ and
 $^A\Delta _{z0}=-\frac{1}{2}(u_0t_z-3u_zt_z-2u_{\perp}t_z)$,\\
 which implies $^A\Delta^{CDW}_{z0}=-\frac{1}{2}(u_0-3u_z-2u_{\perp})$ and
$^A\Delta^{KD}_{x0}=-\frac{1}{2}(u_0-u_z-4u_{\perp})$,\\ \\
$\varepsilon^{CDW}_{\pm\pm}=\pm E_Z\pm 
 \sqrt{E_{kin}(p)\,^2+\big(\,^A\Delta^{CDW}_{z0}\big)^2},\label{eqn:SPL_CDW}$ and  
$\varepsilon^{KD}_{\pm\pm}=\pm E_Z\pm 
 [E_{kin}(p)-\,^A\Delta^{KD}_{x0}] $,\\ \\
 $\Delta\varepsilon_{bulk} =2|E_Z-|^A\Delta _{z0/x0}||.$\\ 
 \hline\hline
\end{tabular}
 \caption{Analytical expressions for the mean-field potential of the symmetry-breaking terms, $^A\,\Delta$, 
the eigenvalues of the full mean-field Hamiltonian, $\varepsilon_{\pm\pm}$, and the minimum gaps in the bulk 
and at the edge, $\Delta\varepsilon_{bulk/edge}$. We denote by $\mathbf{s}$ and $\mathbf{t}$ the SP spin and isospin 
configuration of the two electrons per orbital and $\theta$ describes the canting angle between the two spins. 
In all the formulas for the mean-field potentials, we dropped a constant term $-\frac{1}{2}(u_0+2u_{\perp}+u_z)\mathbb{1}\otimes\mathbb{1}$. These analytical results have been obtained within the approximation that the bulk phase does not change as a function of space when approaching the edge.}
\label{tab:AnalyticExpr}
\end{table}

The ansatz that we introduced in Secs.~\ref{section:TheoFrame} and \ref{section:HF_treat} 
is able to describe spatial dependence of the order and also to capture spin/valley entanglement.

\begin{figure}[t]
  \centering
\includegraphics[width=0.6\textwidth]{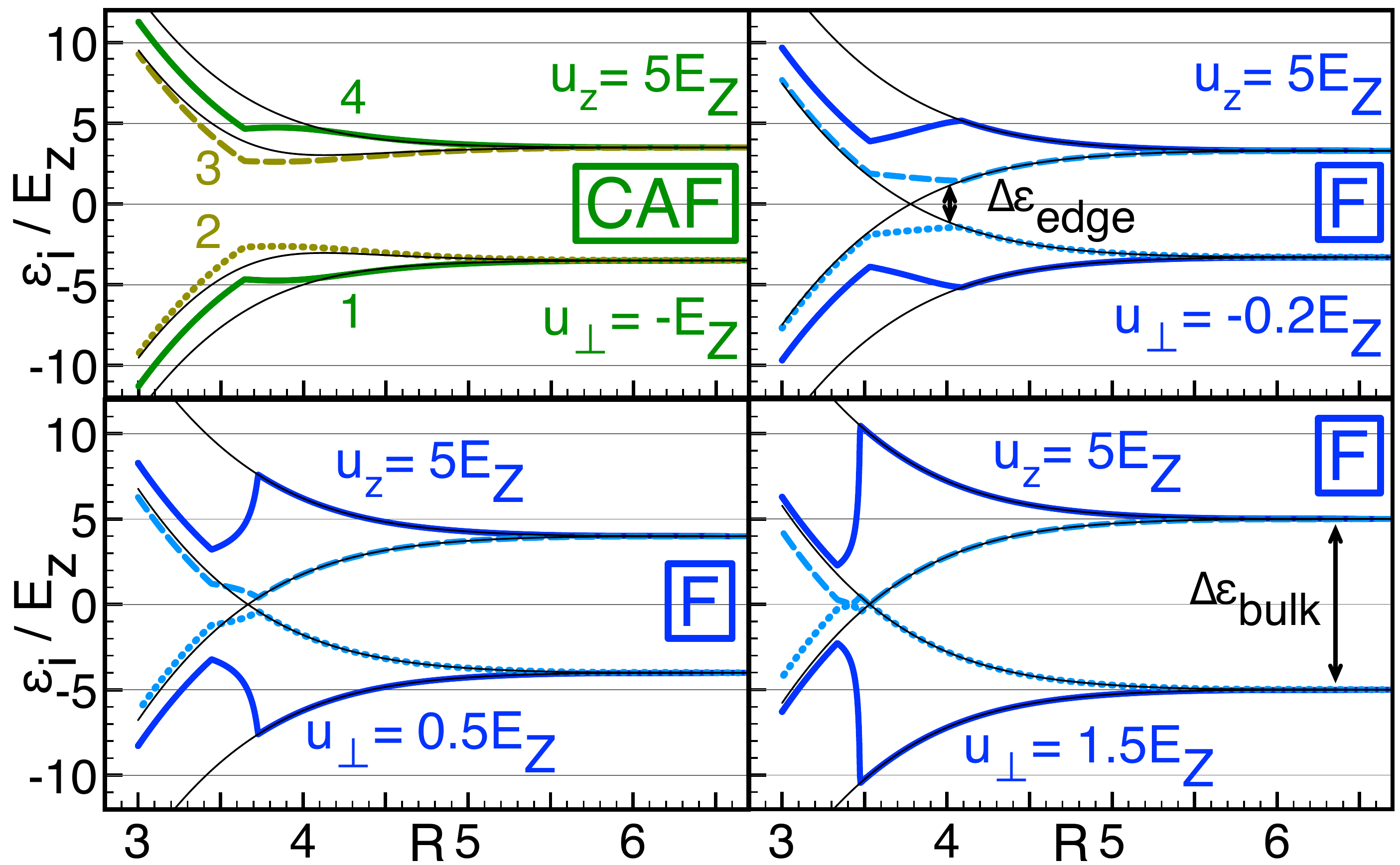}
  \caption{Spatial behavior of the energetic SP spectra in the presence of a 
 boundary for positive transverse coupling  $u_z=5E_Z$ and 
different values of the perpendicular coupling $u_{\perp}$~: CAF bulk 
phase at $u_{\perp}=-E_Z$ in the upper left panel (green lines), F bulk phase at 
$u_{\perp}=-0.2E_Z$, $u_{\perp}=0.5E_Z$, and $u_{\perp}=1.5E_Z$ in the remaining 
panels, respectively (blue lines). Thick, colorful lines show our numerical results. Here, different line shapes distinguish between the 
different single-particle energy levels 
$\varepsilon_1\le\varepsilon_2\le\varepsilon_3\le\varepsilon_4$.  Thin, black 
lines compare to the analytical formulas for $\varepsilon_{\pm\pm}(R)$ listed in Table \ref{tab:AnalyticExpr} 
for the different phases, respectively, 
in which no modulation of the underlying spin/isospin texture towards the edge 
is taken into account.
We see two bulk levels, being two-fold degenerate and separated by a gap of 
width $\Delta \varepsilon_{bulk}$ in the bulk, split into four branches when 
approaching the edge. The branches exhibit kinks and regimes of different 
behavior corresponding to the transitions between different spin and isospin 
phases during the evolution from the bulk to the edge (see 
Fig.~\ref{fig:Obs_AllPhases}). The edge gap between the two intermediate levels 
$\varepsilon_2$ and $\varepsilon_3$ (dotted and dashed lines, respectively) $\Delta \varepsilon_{edge}=\text{min}(|\varepsilon_3-\varepsilon_2|]$ remains 
finite over a certain range of $u_{\perp}$, reducing gradually as $u_{\perp}$ 
increases until it finally closes completely. The lower right panel shows a 
configuration were the levels cross and form gapless edge states.
Special attention should be paid to the upper right panel and the lower left 
panel in which the numerical results show configurations with a F phase in the 
bulk were finite edge gaps $\Delta \varepsilon_{edge}\neq0$ remain, whereas the 
analytical curves cross as soon as the bulk passes into an F phase.
The behavior of the $\Delta \varepsilon_{edge}$ as a function of $u_{\perp}$ is 
studied further in Fig.~\ref{fig:MinGap_vsu}.}
    \label{fig:MFSpectr_AllPhases}
\end{figure}

\begin{figure}[t]
  \centering
\includegraphics[width=0.6\textwidth]{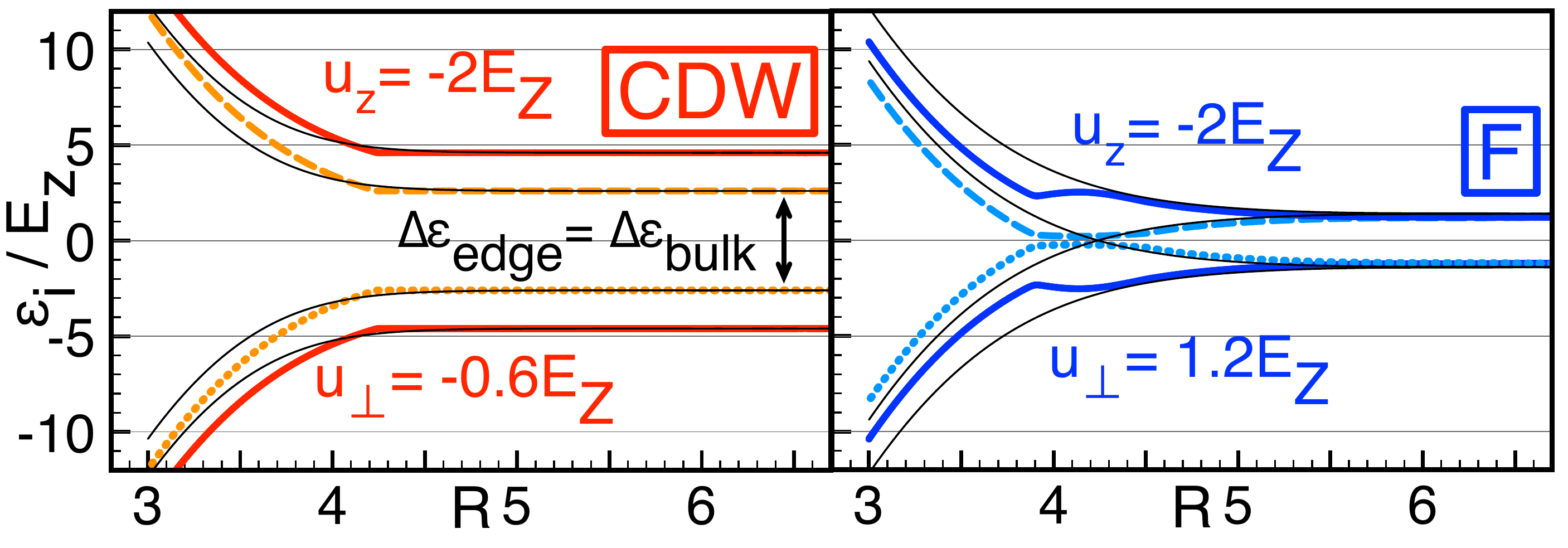}
  \caption{Same as Fig.~\ref{fig:MFSpectr_AllPhases} for $u_{z}=-2E_Z$ and 
different $u_{\perp}$, hence displaying the SP spectra for the bulk phase 
regimes of the lower half plane of the GS phase diagram. Left panel: CDW bulk 
phase at $u_{\perp}=-0.6E_Z$ (red lines). Right panel: F bulk phase at 
$u_{\perp}=-1.3E_Z$ (blue lines). Thin, black lines compare to the analytical 
formulas given in Table \ref{tab:AnalyticExpr} for the different phases, respectively, in 
which the evolution of the bulk's phase towards the edge is neglected.
For the CDW wave phase in the left panel we observe four non-degenerate SP 
levels in the bulk; furthermore, as these levels do not bend towards each other 
but disperse when approaching the edge, the minimum energy gap to SP excitations 
is given by the bulk gap $\Delta \varepsilon_{edge}=\Delta \varepsilon_{bulk}$.}
    \label{fig:MFSpectr_AllPhases_LHS}
\end{figure}

We obtain the set of parameters $\{\alpha(R), \beta(R), \chi(R)\}$ 
characterizing the GS $|g\rangle$, by minimizing the total energy $E_{tot}$ of 
Eq.~(\ref{eqn:Etot}). From these space-dependent parameters, we construct the 
corresponding density matrix $\rho_g=gg^{\dagger}$ via Eq.~(\ref{eqn:rho_SNR}), 
which in turn allows us to reconstruct and diagonalize $\h^{HF}$ of 
Eq.~(\ref{eqn:hHF}). The analysis is repeated for every point in space, thereby 
yielding the spatial behavior as a function of the 
distance from the edge. 
The results for the spectra $\varepsilon_i$ established near the edge for 
different phases in the bulk are shown in Figs.~\ref{fig:MFSpectr_AllPhases} and 
\ref{fig:MFSpectr_AllPhases_LHS}. We have chosen the same parameters  
 as in the  plots of Figs.~\ref{fig:Obs_AllPhases} and \ref{fig:MaxConc_vsu}. 
For $u_z>0$ in
Fig.~\ref{fig:MFSpectr_AllPhases}, we explore the upper half plane, whereas for {$u_z<0$} (Fig.~\ref{fig:MFSpectr_AllPhases_LHS}), the evolution of the bulk 
phases of the lower half-plane is displayed. In all figures the thick
colorful lines show the outcome of our numerical studies. The 
thin black lines represent the analytical results for the SP energy levels for 
the phase established in the bulk at the particular system parameters shown, as 
given in Table \ref{tab:AnalyticExpr}.
Figure \ref{fig:MFSpectr_AllPhases} illustrates the evolution of the edge SP 
levels for a transverse coupling energy of $u_z=5E_Z$ and different values of 
the perpendicular coupling $u_{\perp}$, for which the bulk phase 
configuration passes from a CAF phase at $u_{\perp}=-E_Z$ (green lines, upper 
left panel) to a F phase at $u_{\perp}=-0.2E_Z, u_{\perp}=0.5E_Z$, and 
$u_{\perp}=1.5E_Z$, respectively (blue spectra in the upper right and the two 
lower panels). 
In general, the spectra in all four panels show the following behavior~: two 
flat energy levels, separated by the gap $\Delta \varepsilon_{bulk}$, are 
present in the bulk, both two-fold degenerate and  they split into four 
branches when approaching the edge. The two intermediate levels, being labeled 
$\varepsilon_2$ and $\varepsilon_3$, first bend towards each other, establishing the
minimum energy gap $\Delta \varepsilon_{edge}<\Delta \varepsilon_{bulk}$, 
before, even closer to the edge, the two lowest and the two highest levels 
$\varepsilon_1, \varepsilon_2$ and $\varepsilon_3,\varepsilon_4$ are driven 
apart in two parallel pairs, respectively.

In Fig.~\ref{fig:MFSpectr_AllPhases_LHS}, we show the spectra corresponding to 
the phases of the lower half plane~: at $u_z=-2$, we display the energy levels at 
$u_{\perp}=-0.6E_Z$ (red lines in the left panel), for which the bulk 
is CDW as well as for $u_{\perp}=1.2E_Z$ (blue lines in the 
right panel), where the bulk is F. 
The behavior of the SP energy levels in a CDW bulk phase qualitatively differs 
from the situation of the CAF/F phase described above. In the left panel of 
Fig.~\ref{fig:MFSpectr_AllPhases_LHS} there are four non-degenerate levels in the bulk. 
In 
contrast to the levels of the CAF/F case, they do not bend towards each other  
and there is no minimum energy induced by the edge behavior.
Hence, we find that the minimum edge 
gap is equal to the bulk gap~: $\Delta \varepsilon_{edge}=\Delta 
\varepsilon_{bulk}$. 
Sufficiently close to the edge, the levels again form two parallel pairs.
The spectra for the KD phase in the bulk are not shown because, as mentioned in 
Sec.~\ref{section:PropGS},  this state does not undergo any significant
evolution when approaching the edge. The spectra do not differ from the analytical prediction for 
$\varepsilon_{\pm\pm}$ shown for $^A\Delta_{x0}^{KD}$ in the lower right panel of 
Fig.~\ref{fig:SPL_anltc}.

When comparing this behavior to the analytical results as given in Table \ref{tab:AnalyticExpr} (black lines), deep in the bulk 
we find that all curves coincide as they should. Furthermore, from the discussion 
in Sec.~\ref{sssection:SVent}, we know that there is no spin-valley 
entanglement in the bulk, i.e., the bulk states indeed are of separable 
product form. Yet, significant deviations between the numerical results 
capturing the full GS properties and the analytical curves from the simplified 
treatment are observed when moving closer to the edges, where the GS
spin/isospin configuration starts to deviate from the bulk phase; cf.~Fig.~\ref{fig:Obs_AllPhases}. The SP energies $\varepsilon_i$ have
kinks whenever the underlying spin/isospin texture changes and exhibit qualitatively different behavior in the 
different texture regimes. Thus, the emergence of different spin/isospin 
configurations due to the edge potential when approaching the edges directly 
translates into the SP spectra leading to a complex energy structure as a 
function of space.

\subsection{Single Particle Level Crossings in the different Texture Regimes}
\label{sssec:SPLCrossings_Phases}

\begin{figure}[t]
  \centering
\includegraphics[width=0.6\textwidth]{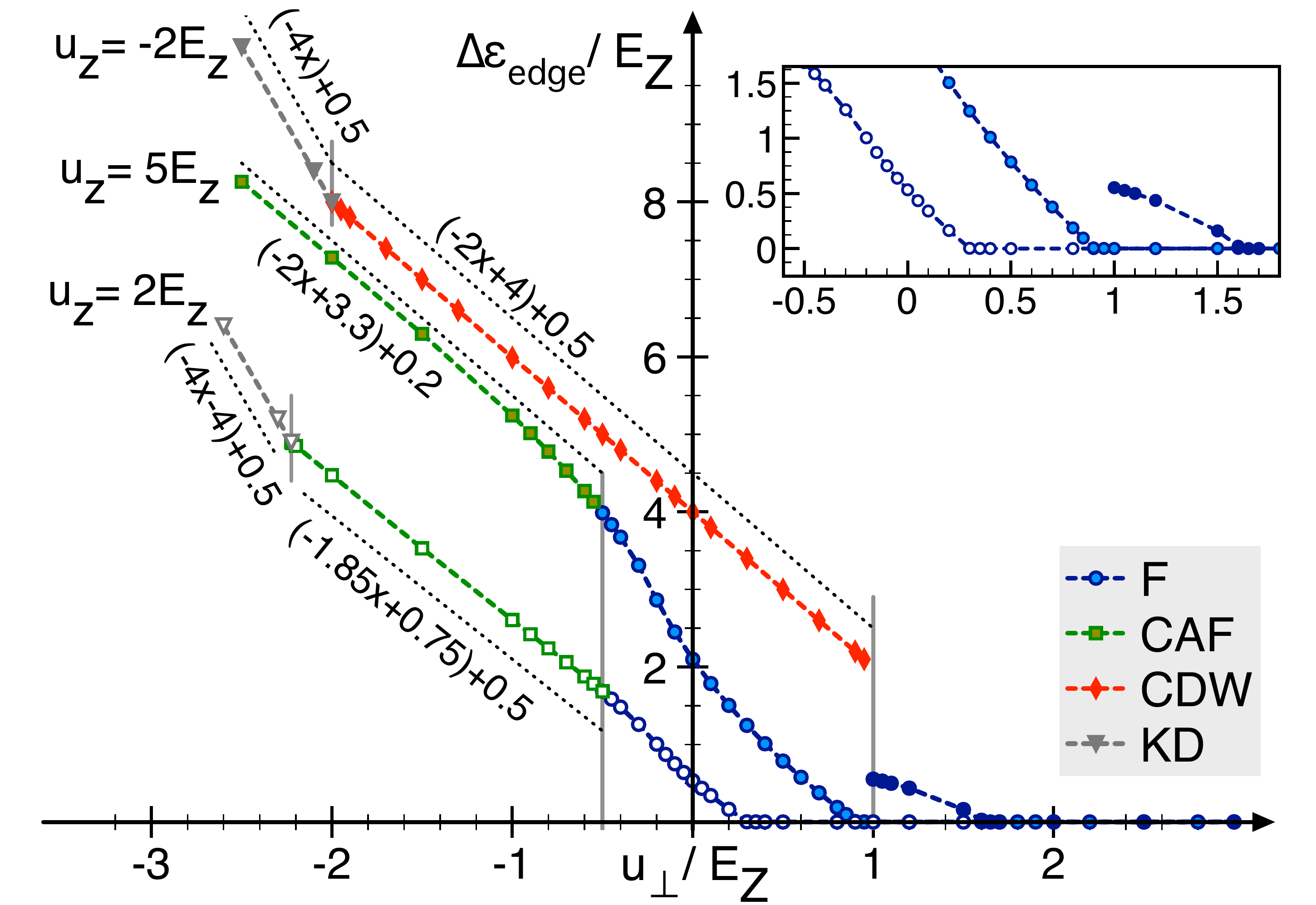}
  \caption{Behavior of the gap $\Delta \varepsilon_{edge}$ in 
the SP spectra close to the edge. We use three
couplings~: $u_z=-2E_Z$, $u_z=2E_Z$, and $u_z=5E_Z$ (filled, shaded and empty 
symbols, respectively). Different colors and 
symbols stand for different bulk phases: KD (grey triangles), CDW (red 
diamonds), CAF (green squares), and F phase (blue circles). Dashed, colored 
lines connect the data points as a guide to the eye. The dotted, black lines 
represent the behavior of the data in the linear regimes (they are shifted by a 
constant offset with respect to the curves for better visibility). Gray
vertical lines indicate the critical values for bulk phase transitions. 
For all phases the gap $\Delta\varepsilon_{edge}$ monotonically decreases as 
$u_{\perp}$ grows until it finally closes in the regime where the bulk is in an F 
phase. The inset is a close-up on how the blue lines smoothly approach 
$\Delta\varepsilon=0$.
The transitions from $\Delta 
\varepsilon_{edge}\neq0$ to $\Delta \varepsilon_{edge}=0$ take place at 
$u_{\perp}\approx\, 0.3E_Z$, $u_{\perp}\approx\, E_Z$, $u_{\perp}\approx\, 
1.625E_Z$. }
    \label{fig:MinGap_vsu}
\end{figure}

\begin{figure}[t]
  \centering
\includegraphics[width=0.6\textwidth]{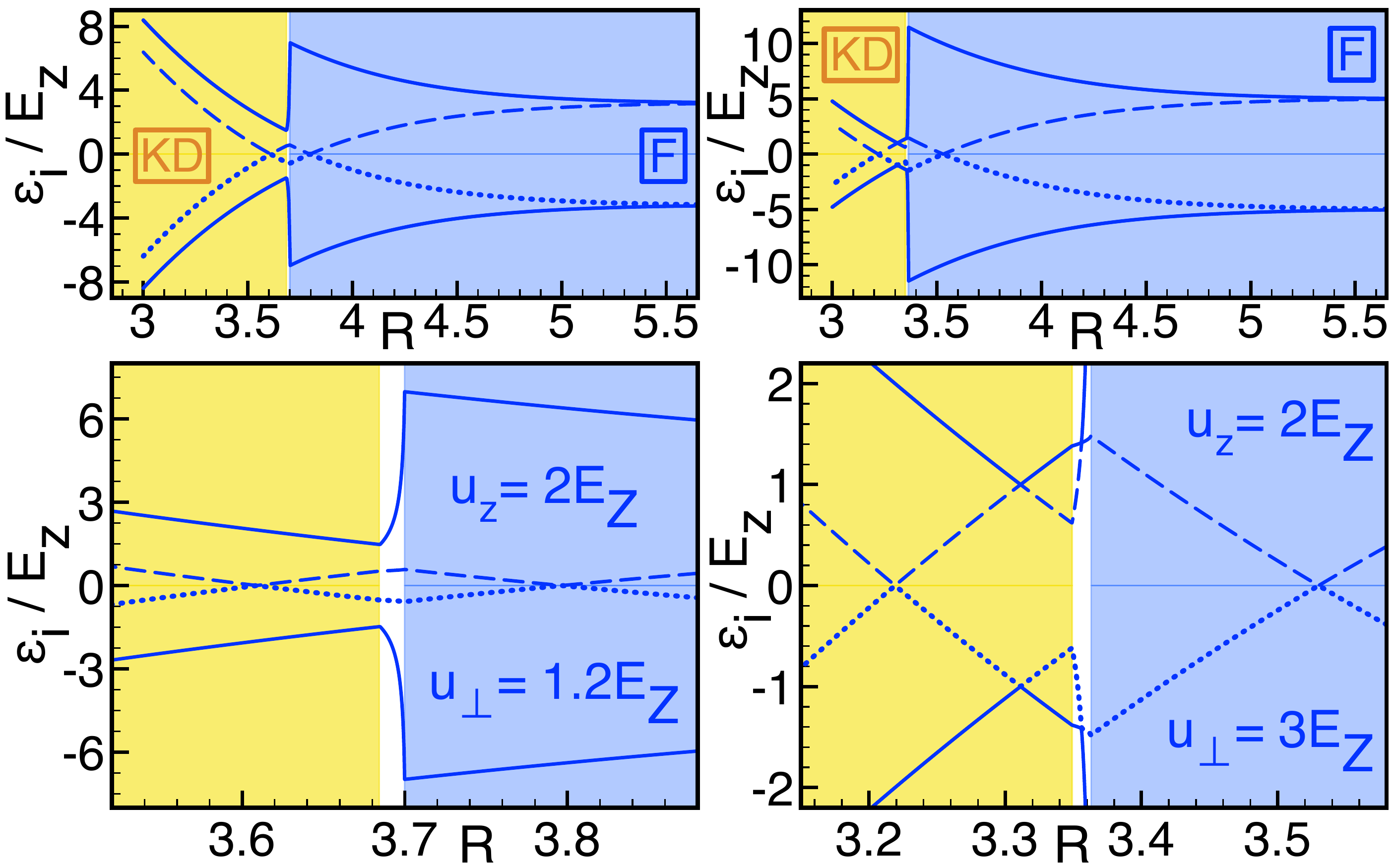}
  \caption{Close-up on the SP spectra in the regime with multiple 
crossings for $u_z=2E_Z$. Left side:  $u_{\perp}=1.2E_Z$, right side: 
$u_{\perp}=3E_Z$. The blue lines show the SP energy levels $\varepsilon_i$. The different background 
colors mark the spatial regimes in which the GS establishes different 
spin/isospin textures: there is 
 a F phase (blue region). In the intermediate (white) region, the 
system undergoes a transition before it finally ends up in a KD phase 
(yellow region).
We observe the crossings of the SP levels to occur in regions of different 
spin/isospin phases - their origins lie in the different symmetries of the F and 
KD phases.}
    \label{fig:MFSpectr_Cross}
\end{figure}

The energy levels 
of the SP ground and excited states show a complex structure as a function of 
space depending on the spatial changes of the spin and isospin texture when 
approaching the graphene edge. In particular, in some configurations, the SP 
spectra exhibit a finite gap, whereas for other system parameters, the edge 
states are gapless when the SP levels  cross. In this section, we 
investigate  the crossings between SP energy levels which leads to
gapless edge excitations. We first discuss the properties of the edge 
gap and its behavior when approaching the critical values where it closes. Then, 
we explain the  number of allowed crossing points between 
energy levels and the connection to the symmetry properties of the underlying 
spin/isospin texture phases.
The spatial variation of the order parameters has a direct impact on the overall shape
of the dispersion of edge modes. This is readily seen in 
Figs.~\ref{fig:MFSpectr_AllPhases} and
\ref{fig:MFSpectr_AllPhases_LHS}, where we plot the dispersions from our calculation
including edge effects and results from a similar calculation using only bulk values without spatial variation.
In order to investigate the closure of the edge gap $\Delta \varepsilon_{edge}$ as 
a function of the ratio $u_{\perp}/E_Z$, we evaluate the size of the minimum gap 
in the SP spectra for various system parameters. 
We choose the same values for the anisotropy energies 
as in Sec.~\ref{section:PropGS} by fixing $u_z=5E_Z, u_z=2E_Z$ and 
$u_z=-2E_Z$ and varying $u_{\perp}$
so that we access all bulk phases in the 
GS phase diagram. The resulting 
curves $\Delta \varepsilon_{edge}({u_{\perp}}/{E_Z})$ are shown in 
Fig.~\ref{fig:MinGap_vsu}. We find that the size of the edge gap $\Delta 
\varepsilon_{edge}$ is a strictly monotonic decreasing function of 
${u_{\perp}}/{E_Z}$ for all values of $u_z$. 
When the bulk is KD or CDW  the flat bulk SP levels split 
further apart when approaching the edge 
so that the minimum gap in the spectrum is equal to the bulk gap 
$\Delta\varepsilon_{edge}=\Delta\varepsilon_{bulk}$. In these two 
cases we find that the bulk gap is a linear function of the perpendicular 
coupling energy~: $\Delta\varepsilon_{bulk}^{KD/CDW}\propto u_{\perp}/E_Z$. The 
numerical results in Fig.~\ref{fig:MinGap_vsu} follow exactly the analytical 
prediction given in Table \ref{tab:AnalyticExpr}~: At $u_z=-2E_Z$, we find 
$\Delta\varepsilon^{KD}_{bulk}=-4u_{\perp}$ and 
$\Delta\varepsilon^{CDW}_{bulk}=-2u_{\perp}+4E_Z$, whereas the KD edge gap at 
$u_z=2E_Z$ behaves as $\Delta\varepsilon^{KD}_{bulk}=-4u_{\perp}-4E_Z$. These 
analytical curves are plotted in Fig.~\ref{fig:MinGap_vsu} as dotted lines for 
comparison (they are shifted by a constant offset with respect to the numerical 
results for better visibility). 
As a 
consequence of this linear behavior in $u_\perp$, for 
 couplings favoring KD or CDW in the bulk, at the system 
parameters chosen in Fig.~\ref{fig:MinGap_vsu}, there is always a non-zero gap 
to SP excitations.
When the GS in the bulk is a 
CAF or a F phase, the SP spectra now bend towards each other and 
therefore exhibit a minimum energy gap $\Delta\varepsilon_{edge}$ near the edge 
which is smaller than the bulk gap.

For $u_{\perp}\le-E_Z/2$ where the bulk is a CAF phase (green 
squares), the spectrum always exhibits an 
non-zero edge gap which is almost linear as a function of the perpendicular 
coupling. 
At values $u_{\perp}\ge-E_Z/2$, hence for a F bulk, 
the shape of the spectrum changes qualitatively and the bulk gap closes 
 in a non-linear way, asymptotically approaching zero at sufficiently large values of $u_{\perp}$.
For the transverse couplings chosen in Fig.~\ref{fig:MinGap_vsu}, 
$u_{z}=5\,E_Z$, $u_{z}=2\,E_Z$, and $u_{z}=-2\,E_Z$, the edge gap $\Delta
\varepsilon_{edge}$ closes at $u_{\perp}\approx  \, E_Z$, $u_{\perp}\approx  \, 
0.3E_Z$, and $u_{\perp}\approx  \, 1.6E_Z$, respectively. 
All these values leads to a bulk which is ferromagnetic. So the prediction for 
the gap closure point clearly differs from the value $u_{\perp}=-\frac{E_Z}{2}$, which can be read 
from the bulk phase diagram in Fig.~\ref{fig:PhaseDiag_Khar}.
This is due to the changes of the spin/isospin configuration of the GS 
induced by the effective edge potential as we approach the boundary; cf.~Fig.~\ref{fig:Obs_AllPhases} of Sec.~\ref{section:PropGS}. Indeed the system does 
not remain in a F phase configuration all the way from the bulk to the edge. 
During its transition into a KD phase close to the boundary, there is
an intermediate regime with non-zero 
spin-valley entanglement and simultaneous canting of both spin and isospin. 
Hence, in this transition regime there is  no \textit{a priori} justification for $\Delta_{edge}^{CAF/F}$ of 
Table \ref{tab:AnalyticExpr} to yield a correct description of the edge gap.

From this analysis of the gaps of the SP spectra we can draw the following conclusion~:
when the bulk is  CDW,  KD or CAF phase, the SP energy levels 
always have non-zero gaps. However
for a bulk F phase one may have gapped or gapless spectra.
We note that ignoring the spatial variation of the trial HF state leads to 
qualitatively different results~\cite{kharitonov_edge_2012}.


\subsection{Number of level crossings}

We now discuss in more detail the number of crossings of the HF single-particle states.
In the F phase with dispersion $\varepsilon_{\pm\pm}^{F}$ in Table \ref{tab:AnalyticExpr}, the 
intersecting levels  $\varepsilon_{+-}^{F}$ and 
$\varepsilon_{--}^{F}$  cross exactly once as their slope is given 
by the slope of the kinetic energy term  
and they are monotonic functions of the spatial 
coordinate. 
For several system parameters, such as in the spectrum 
in the lower left panel of Fig.~\ref{fig:MFSpectr_AllPhases} at $u_z=5E_Z, 
u_{\perp}=1.5E_Z$ as well as in close ups  shown in 
Fig.~\ref{fig:MFSpectr_Cross}, we observe  multiple 
crossings. 
We first discuss the occurrence of multiple 
crossings and the relation with the underlying spin/valley texture. 
The number of crossings is governed by the symmetries of the HF Hamiltonian
and the magnitude of the HF self-consistent potentials. 
After discussing 
the different phases separately, we apply the insights to the edge state 
structure described in Sec.~\ref{section:PropGS}, where the GS phase changes 
as a function of space when approaching the edge from the bulk.
We first discuss the case of the CAF/F transition. We rewrite the mean-field 
Hamiltonian of Eq.~(\ref{eqn:hHF}) involving the CAF/F mean-field potential given in Table \ref{tab:AnalyticExpr} 
by decomposing it into four $2\times2$ matrices as~:
\begin{equation}
\h^{HF}(p)=
\begin{bmatrix}
 \gamma_1 & \gamma_2\\
 \gamma_3 & \gamma_4
\end{bmatrix},
\label{eqn:hHF_Matrix}
\end{equation}
where the respective entries for the CAF/F phase are given by~:
\begin{align}
\nonumber\gamma _1&=\,^A\Delta_{zx} \sigma^{}_x-(E_Z-\,
^A\Delta _{0z})\,\sigma^{}_z,\\
\nonumber\gamma _2&=\gamma _3=-E_{kin}(p)\;\mathbb{1}^{},
\\
\gamma _3&=-\,^A\Delta_{zx} \,\sigma^{}_x-(E_Z-\,^A\Delta _{0z})\,\sigma^{}_z,
\end{align}
with 
$\,^A\Delta _{zx}$ and $\,^A\Delta_{0z} $ defined for the CAF/F phase in Table \ref{tab:AnalyticExpr}. 
The size of the gap is therefore governed by the first off-diagonal coupling 
matrix elements $\,^A\Delta^{CAF/F}_{zx}$. If $\,^A\Delta^{CAF/F}_{zx}\neq0$, as 
is the case for any non-zero canting angle $\theta\neq0$, the eigenvalues of 
the Hamiltonian $\h^{HF}_{CAF/F}$ exhibit the characteristic behavior of 
avoided crossings. The SP levels 
are allowed to cross only for $\,^A\Delta^{CAF/F}_{zx}=0$ at $\theta=0$, i.e., in the F phase.
In the bulk, i.e., at $E_{kin}\equiv0$, all values of the coupling 
energies $u_{z}$ and $u_{\perp}$ allowed for the F phase yield the same ordering 
of the  SP energy levels 
$\varepsilon_{\pm\pm}^{F,0}=\varepsilon_{\pm\pm}^{F,0}(E_{
kin}\equiv0)$, independently of the sign or the modulus of 
$\Delta_{0z}^{CAF/F}$: 
$\varepsilon_{+-}^{F,0}=\varepsilon_{-+}^{F,0}<0<\varepsilon_{--}^{F,0}
=\varepsilon_{++}^{F,0}$. 
Hence, there is only one possible scenario of level 
crossings when approaching the boundary as the increasing edge potential is 
driving the SP levels away from their bulk values. This leads to exactly one 
crossing of the levels $\varepsilon_{+-}^{F}$ and $\varepsilon_{--}^{F}$, 
shown by the blue, dashed lines in the upper left panel of 
Fig.~\ref{fig:SPL_anltc}.
We can perform the same analysis for  CDW or  KD phases. Again, 
we rewrite the corresponding HF Hamiltonians of Eq.~(\ref{eqn:hHF}) with the 
potentials $\,^A\Delta^{CDW/KD}_{z0/x0}$ from Table \ref{tab:AnalyticExpr} and 
 we find the 
respective entries for the CDW~:
\begin{align}
\nonumber\gamma _1&=-E_Z\,\sigma^{}_z+\,^A\Delta _{z0}\,
\mathbb{1}^{},\\
\nonumber\gamma^{ }_2&=\gamma^{ }_3=-E_{kin}(p)\,\mathbb{1}^{},\\
\gamma^{ }_4&=-E_Z\,\sigma^{}_z-\,^A\Delta^{ }_{z0}\,\mathbb{1}^{},
\end{align}
whereas for the KD phase we find~:
\begin{align}
\nonumber\gamma^{ }_1&=\gamma^{ }_4=E_Z\,\sigma^{}_z,\\
\gamma^{ }_2&=\gamma^{ }_3=\Big(\,^A\Delta^{KD}_{x0}-E_{kin}(p)\Big)\,\mathbb{
1}^{}.
\end{align}
The Hamiltonians for the CDW phase and the KD phase thus turn out to have 
higher symmetry than in the CAF phase: In $\h^{HF}_{CDW}$ and $\h^{HF}_{KD}$, 
all entries of the two first off-diagonals as well as of the antidiagonal are 
zero. Pairwise degeneracy of the corresponding eigenvalues, i.e., crossings between the SP energy levels, is now allowed. 
Note that, unlike the transition from a CAF to a F phase, all other transitions 
 do not correspond to smooth transitions. In these cases, a
transition between phases goes along with an abrupt change of the 
symmetry properties of the spin/isospin configuration of the GS and the 
corresponding Hamiltonian. 

We now discuss the different possible scenarios of SP level crossings in  CDW 
 and   KD phases.
The SP energy levels of the CDW $\varepsilon_{\pm\pm}^{CDW}$ 
 in Table \ref{tab:AnalyticExpr} are independent of the sign of 
$\,^A\Delta^{CDW}_{z0}$. Different orderings of the bulk levels 
$\varepsilon_{\pm\pm}^{CDW,0}$ at $E_{kin}\equiv0$ may, however,
appear depending on the modulus of $\,^A\Delta^{CDW}_{z0}$~: for 
$|\,^A\Delta^{CDW}_{z0}|>E_Z$, the bulk states are ordered as  
$\varepsilon_{--}^{CDW,0}<\varepsilon_{+-}^{CDW,0}<\varepsilon_{-+}^{CDW,0}
<\varepsilon_{++}^{CDW,0}$. 
In this case, when approaching the boundary, the 
kinetic energy potential drives the positive and the negative energy states 
further apart from each other such that they do not cross. In the case where 
$|\,^A\Delta^{CDW}_{z0}|<E_Z$, however, the bulk states rather follow the 
hierarchy 
$\varepsilon_{--}^{CDW,0}<\varepsilon_{-+}^{CDW,0}<0<\varepsilon_{+-}^{CDW,0}
<\varepsilon_{++}^{CDW,0}$.
In this case, turning on the effective edge 
potential drives the levels $\varepsilon_{-+}^{CDW,0}$ and $\varepsilon_{
+-}^{CDW,0}$ \textit{towards} each other and they cross at zero energy. These 
two different scenarios are depicted in the upper right panel of 
Fig.~\ref{fig:SPL_anltc}, where the red, solid lines show the levels 
$\varepsilon_{\pm\pm}^{CDW}$ from Table \ref{tab:AnalyticExpr} at 
$\,^A\Delta^{CDW}_{z0}=2E_Z>E_Z$ and the black, dashed lines show the spectrum 
for $\,^A\Delta^{CDW}_{z0}=0.3E_Z<E_Z$. The latter case,
$|\,^A\Delta^{CDW}_{z0}|<E_Z$, however, is prohibited by the conditions imposed on the couplings
 $u_z$ and $u_{\perp}$ in 
order for the system to establish a CDW phase in the bulk. Requiring 
$u_z<u_{\perp}$ and $u_z<-E_Z-u_{\perp}$ will always force 
$|\,^A\Delta^{CDW}_{z0}|>E_Z$. Therefore, treating the system as having a 
stable CDW phase in the bulk and all the way to the edge will never lead to any 
crossings of the SP edge levels.
Turning to the more important case of the KD phase, the situation becomes even 
richer. Here, depending on the sign and the modulus of $^A\Delta^{KD}_{x0}$, 
four different SP level orderings in the bulk and four resulting crossing 
scenarios may appear. For $|^A\Delta^{KD}_{x0}|<E_Z$, if $^A\Delta^{KD}_{x0}>0$, 
there is one level crossing at zero energy and two additional crossings above 
and below the zero energy line, respectively, whereas for negative 
$^A\Delta^{KD}_{x0}$, only one crossing at zero energy is present. The case 
$|^A\Delta^{KD}_{x0}|>E_Z$ can lead to four crossings, two at zero energy plus 
one above and one below, respectively, if $\Delta^{KD}_{x0}>0$, whereas for 
$^A\Delta^{KD}_{x0}<0$, the four levels do not cross.

Examples of the four different cases are shown in the lower panels of 
Fig.~\ref{fig:SPL_anltc}, where the lower left panel shows the possible 
situations for $|\,^A\Delta^{KD}_{x0}|<E_Z$ (solid, orange lines for 
$\,^A\Delta^{KD}_{x0}=-0.3E_Z<0$ and black dashed lines for 
$\,^A\Delta^{KD}_{x0}=0.7E_Z>0$), whereas the right panel displays the 
corresponding spectra for $|\,^A\Delta^{KD}_{x0}|>E_Z$ (here, the solid, orange 
lines are for $\,^A\Delta^{KD}_{x0}=-1.5E_Z<0$ and black dashed lines for 
$\,^A\Delta^{KD}_{x0}=2E_Z>0$). Note that, just as in the case of the 
CDW phase, not all these cases are allowed by the restrictions on the 
parameter range for a KD bulk~: requiring 
the couplings  $u_{z}$ and $u_{\perp}$ to fulfill the relations 
$u_{\perp}<u_z$ and $u_z<\frac{E_Z^2}{2u_{\perp}}-u_{\perp}$ always implies 
$\,^A\Delta^{KD}_{x0}>E_Z$. Therefore, again, all cases including possible 
crossings between edge levels are ruled out for a system with a KD 
phase in the bulk.

This simple picture drawn for constant order parameters
changes when considering the electronic GS structure described in 
Sec.~\ref{section:PropGS}. Indeed the GS spin/isospin texture 
deviates from the bulk phase when moving towards the edge as a consequence of the 
growing edge potential. Close enough to the edge
 the system is always driven into a KD phase. Hence when 
moving sufficiently close to the edge
 the GS does becomes KD-ordered, even though the system 
parameters $u_{z}$ and $u_{\perp}$ do not allow KD order in the bulk. 
Two examples are shown in
the close-ups  in Fig.~\ref{fig:MFSpectr_Cross}. Parameters 
 in both panels are chosen such that the bulk system at 
$E_{kin}\equiv0$ is in a F phase. When moving towards the edge
 the 
energy levels evolve according to $\varepsilon_{\pm\pm}^{F}$ of Table \ref{tab:AnalyticExpr} 
(corresponding to the evolution within the blue 
region). 
A first crossing between the intermediate levels occurs, as 
predicted by the analysis of the F phase energy levels. 
After the transition 
region (left white), the GS becomes KD  (marked by the yellow 
shading). 
However, the system parameters  do 
not force $\,^A\Delta^{KD}_{x0}>E_Z$~: in the left panel of 
Fig.~\ref{fig:MFSpectr_Cross}, we find $\,^A\Delta^{KD}_{x0}=-3.4E_Z$ and in 
the right panel we have $\,^A\Delta^{KD}_{x0}=-7E_Z$. 
Therefore the energy levels now evolve according to $\varepsilon_{\pm\pm}^{KD}$ of 
Table \ref{tab:AnalyticExpr} in 
the case $^A\Delta^{KD}_{x0}<0,\; |\,^A\Delta^{KD}_{x0}|<E_Z$. 
As a consequence, 
in this regime, one more level crossing may occur.
Hence, the appearance of several crossings of the SP energy levels in the 
numerical spectra as in the lower right panel of 
Fig.~\ref{fig:MFSpectr_AllPhases} and in Fig.~\ref{fig:MFSpectr_Cross} can be 
explained combining the insight of Sec.~\ref{section:PropGS} that any bulk 
phase by the edge potential always is driven into a KD phase close to the 
boundary, with the understanding of the  possible behavior of 
$\varepsilon_{\pm\pm}^{KD}$ depending on the value of 
$\,^A\Delta^{KD}_{x0}$ as a function o the coupling energies $u_{z}$ and 
$u_{\perp}$. 
The SP energy levels describing the numerical results of 
Figs.~\ref{fig:MFSpectr_AllPhases}, \ref{fig:MFSpectr_AllPhases_LHS}, and 
\ref{fig:MFSpectr_Cross} can be summarized as~:
\begin{equation}
\varepsilon_{\pm\pm}(R)=\begin{cases}
\varepsilon_{\pm\pm}^{bulk}(R)  & \text{for } R> R_{2},\\
\text{unknown }&\text{for } R_1<R<R_{2},\\
\varepsilon_{\pm\pm}^{KD}(R) & \text{for  }R< R_1,
\end{cases}
\end{equation}
where $\varepsilon_{\pm\pm}^{bulk}(R)$ denotes the level spectra for 
the  bulk phase established at a given choice of system parameters 
and $R_2$ and $R_1$ label the inner and outer limits in space of the domain 
wall, for which there is no simple analytic expression. The 
evolution of $\varepsilon_{\pm\pm}^{KD}(R)$ is no longer limited to 
the non-crossing behavior imposed for a bulk KD phase but it can exhibit 
any of the shapes drawn in the two lower panels of Fig.~\ref{fig:SPL_anltc}. 
Which of these curves describes the KD-like evolution of the edge states 
correctly is determined by the system parameters $u_{\perp}$ and $u_z$ that 
govern the \textit{bulk} texture phase.
From the analysis of the number of SP level crossings we hence learn that, in 
principle, by choosing appropriate values of $u_{z}$ and $u_{\perp}$, SP energy levels
can have zero, one, two, or even three crossings 
 at zero energy. Among these crossings, one is due to the symmetry 
properties of the bulk F phase. The remaining crossings  appear in the 
KD phase close to the boundary, which in this regime shows novel properties not 
present for a KD phase in the bulk. 
We note that these crossings occur at different distances 
from the edge - at the distance where the corresponding KD phase SP levels for a 
certain $\,^A\Delta^{KD}_{x0}$ cross, it is necessary for the system 
already to have evolved from the bulk phase into the KD edge phase in order for 
the additional SP level crossings to occur. This is the reason why 
we do not see any crossings in the SP spectrum shown in the lower left panel of 
Fig.~\ref{fig:MFSpectr_AllPhases_LHS} where the bulk is in a CDW phase~: at the 
distance $R_{cross}$ where the KD-like levels near the edge would cross, the 
system still behaves according to its bulk CDW phase. In this case, the crossing 
is thus prevented by the fact that the crossing point lies outside the KD 
region~: $R_{cross}>R_{2}$. Nevertheless, a situation in which the bulk 
is a CDW but the SP edge states are gapless due to crossings of the 
KD-like levels close to the boundary is not forbidden by the underlying 
symmetry 
principles as the restrictions for the coupling energies of the CDW bulk phase  allow negative 
values of $^A\Delta_{x0}^{KD}$. The exact distances from the edge $R_{1}, R_2$, or 
$R_{cross}$ which define the  points of 
crossing, involve the explicit form of the kinetic energy $E_{kin}(R)$ as they 
are determined by the eventual dominance of the kinetic energy.
Numerical values for 
$R_{1}, R_2$, or $R_{cross}$ therefore strongly depend on the model potential 
chosen for $E_{kin}(R)$. This is not true, however, for the answer to the 
question of whether crossings  are allowed or not
 since the values of $^A\Delta$ are determined generically by the system 
parameters $u_{\perp}$ and $u_{z}$.

\subsection{Properties of the underlying SP States}

\begin{figure}[t]
  \centering
\includegraphics[width=0.6\textwidth]{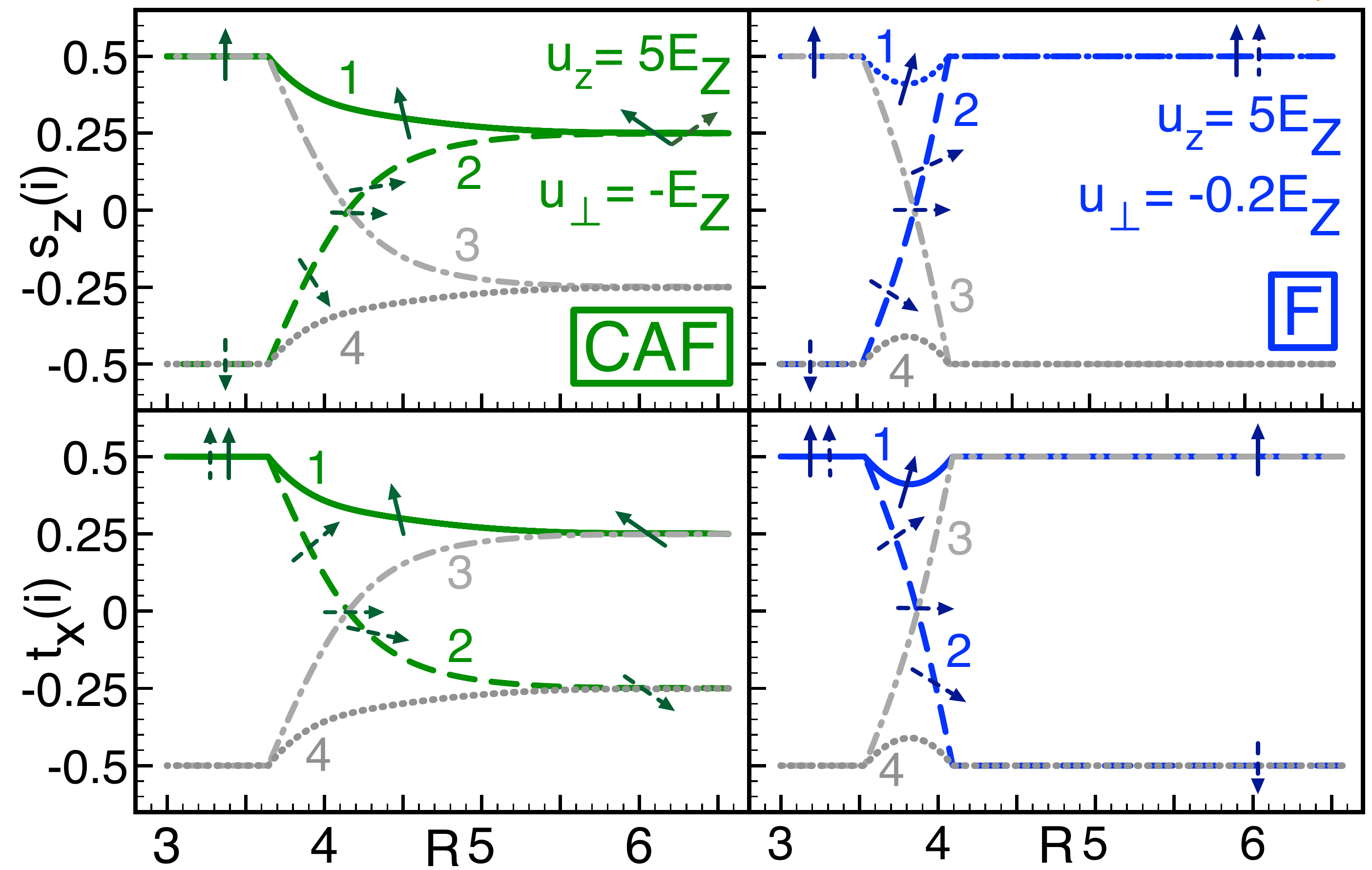}
  \caption{Evolution of the single-electron spin and isospin components 
$s_{z}(i)$ and $t_x(i)$ of the SP eigenstates   
 from the bulk towards the edge for the 
two anisotropy energies $u_{\perp}= -E_Z$ (CAF bulk GS , left panels with green lines) and $u_{\perp}=-0.2\,E_Z$ (F 
bulk GS, right panels with blue lines). Different line shapes distinguish 
between the four SP energy levels 
$\varepsilon_1\le\varepsilon_2\le\varepsilon_3\le\varepsilon_4$. Green/blue 
lines correspond to the observables of the two lowest-lying states 
which are occupied orbitals in the HF GS. Gray lines indicate 
the behavior of the higher-lying SP states. 
Arrows show the behavior of the spin and isospin 
polarizations. The second 
and  third levels $|2\rangle$ and $|3\rangle$ are
oppositely polarized in spin and isospin at the edge.
In all plots we set $u_z=5\,E_Z$. }
    \label{fig:MFSpectr_SPSates_AllPhases}
\end{figure}

In order to obtain a better understanding of the nature of the excited states 
we analyze the properties of the single-electron states.
We compute the spin and isospin components $s_z(i)=\frac{1}{2}\langle 
i|\sigma_z|i\rangle$ and $t_x(i)=\frac{1}{2}\langle i|\tau_x|i\rangle$ and display
the results as a function of space in Fig.~\ref{fig:MFSpectr_SPSates_AllPhases}.

The evolution of the observables  can be summarized as follows~:

\begin{equation}
\begin{array}{cccc|cccc}
 s_z(i)  & \text{edge}  &  \text{intermediate} &  \text{bulk} \hskip10pt & t_x(i)  
&\hskip5pt \text{edge}  &  \text{intermediate} &  \text{bulk}  \\
 \hline
 s_z(1):  & \uparrow  & \nearrow & \uparrow &  t_x(1):  & \uparrow  & \nearrow & \uparrow\\
 s_z(2):  & \downarrow  & \searrow\longrightarrow \nearrow & \uparrow &  t_x(2):  & \uparrow  
& \nearrow\longrightarrow \searrow & \downarrow\\
 s_z(3): &\uparrow&\nearrow\longrightarrow \searrow &\downarrow &  t_x(3): &\downarrow  
& \searrow\longrightarrow \nearrow & \uparrow\\
 s_z(4): & \downarrow &\swarrow&\downarrow &  t_x(4): & \downarrow &\swarrow&\downarrow
\end{array}
\label{eqn:TabProp_SPStates}
\end{equation}
where arrows represent schematically the spin and isospin 
vectors.

The HF GS $|G\rangle$ is built from Slater determinants of $|1\rangle$ and 
$|2\rangle$ as the eigenstates of the two lowest lying branches $\varepsilon_1$ 
and $\varepsilon_2$ in Fig.~\ref{fig:MFSpectr_AllPhases}. 
The lowest energy excitations 
correspond to  single-particle excitation from the second to the third level 
$\varepsilon_2 \rightarrow\varepsilon_3$. 
These two states have oppositely polarized spin and isospin components.
The closing of the gap 
$\Delta\varepsilon_{edge}$ between the second and the third SP level in 
Fig.~\ref{fig:MFSpectr_SPSates_AllPhases} hence is a transition from 
insulating to conducting behavior with the counterpropagating 
current-carrying edge states exhibiting opposite spin and isospin polarizations.
This is the behavior of gapless helical edge states of a QSH state~\cite{hasan_textitcolloquium_2010}.

\section{Conclusion}
\label{section:Discussion}

The SU(4) QH ferromagnetism leads to a highly degenerate manifold of ground states
for neutral graphene. This degeneracy is lifted by small lattice-scale anisotropies and 
there is a competition between phases with several types of order. This competition
is affected notably by the substrate supporting the graphene sample. We have used
a simple model of these anisotropies to study the edge properties of neutral graphene
by means of a HF approach. The sharp atomic edge is then described by an effective field
in valley space which modifies the competition between phases. 
Ultimately, whatever the bulk order,
the system is in a KD phase close enough to the edge. In the transition region between this edge order
and the bulk order, we have obtained evidence for an intermediate regime with spin/valley entanglement.
In this regime there is a nontrivial change of the single-particle spectrum. 
We find that the number of levels crossing the Fermi energy can be varied by changing
the parameter $u_{\perp}/E_Z$. This means that there are metal-insulator transitions
when tilting the magnetic field. This is consistent with the experimental findings
of Young \textit{et al.}~\cite{young_tunable_2014}. If we adopt the estimates for the approximate magnetic field dependencies of $E_Z$ and $u_{\perp}$ stated in Refs.~\onlinecite{kharitonov_phase_2012} and \onlinecite{kharitonov_edge_2012} 
as $E_Z(B)\approx 0.7 $B [T]K and  $u_{\perp}(B_{\perp})\approx1-10B_{\perp}$[T]K, 
where $B$ denotes the total magnetic field and $B_{\perp}$ its component perpendicular to the device plane, the values 
for the parameters stated in Ref.~\onlinecite{young_tunable_2014} suggest that the 
authors were able to experimentally tune the ratio $u_{\perp}/ E_Z$ roughly in a range from  -13 to -0.5. The picture we obtain is more complex than that obtained by assuming that the order
does not persist up to the edge~\cite{kharitonov_edge_2012}. Notably the occurrence
of the metal-insulator transition, while it sets constraints on the microscopic parameters,
does not imply that the bulk is CAF ordered.
The observation of a conductance $G\approx2 e^2/h$ which corresponds to two 
conducting channels, i.e., to one single level crossing,
has two possible explanations~: either the bulk is in a F phase leading 
to one crossing unaffected by the KD edge regime, or 
 the bulk has noncrossing SP levels, but the crossing occurs in the KD
regime close to the edge.
Furthermore, our results suggest that the observation of exactly one crossing 
only corresponds to a limited parameter range. 
Varying the anisotropy 
parameters may lead to the observation of conductance values of higher 
multiples of two, corresponding to several crossings in the SP edge spectrum.

Of course there are obvious limitations of our theoretical approach~: we apply a perturbative treatment of the edge, whose validity is limited to 
a certain range [cf.~the discussion following Eq.~(\ref{eqn:Hkin})].  
The appearance of a KD phase in the vicinity of the edge is a direct 
consequence of treating the effective edge potential perturbatively. Furthermore, in 
our calculations we assume the anisotropy energies $u_{\perp}$ and $u_z$ to 
remain constant at their bulk values [as we discuss after introducing 
Eq.~(\ref{eqn:HAniso})]. This approximation certainly becomes less exact 
as we approach the boundary. Also we have neglected the exchange energy effects
that will create textures in the charge-carrying states.

In conclusion, in this paper we have studied the influence of an edge on the $\nu=0$ 
QH state in monolayer graphene. We found that the effective edge potential  induces 
a change of the GS spin/isospin texture. During this 
evolution, novel  phases are observed, involving simultaneous canting of spin 
and isospin as well as non-zero spin/valley entanglement. Phases of this kind 
are not present in the bulk. Furthermore, we analyzed how this spatial evolution 
changes the SP excited states. Here we have shown 
that, as a consequence of the spatial modulation of the underlying spin/isospin 
texture, the direct correspondence between the conductance properties of the 
edge states and the bulk phase is lost. 
The transport properties are governed by either zero, one, or 
multiple SP level crossings. The analysis of the SP 
eigenstates shows that the lowest SP excitation describes counterpropagating helical edge states carrying 
 opposite spin and isospin polarizations.

\begin{acknowledgments}
We acknowledge discussions with Allan H. MacDonald, Inti Sodemann, Feng-Cheng Wu and Ren\'e Cot\'e. 
A.K. gratefully acknowledges support from the German Academic Exchange Service and 
the German National Academic Foundation.
\end{acknowledgments}



\end{document}